\documentclass[twocolumn]{aastex62}
\hypersetup{linkcolor=red,urlcolor=magenta}

\newcommand\aastex{AAS\TeX}

\newcommand{\abund}[1]{$\log N({\rm #1})/N({\rm H})$}
 
\newcommand{\eg}{e.g.}

\newcommand{\figref}[1]{Fig.~\ref{#1}}

\newcommand{\hone}{\ion{H}{1}}

\newcommand{\heone}{\ion{He}{1}}
\newcommand{\hetwo}{\ion{He}{2}}
\newcommand{\cone}{\ion{C}{1}}
\newcommand{\ctwo}{\ion{C}{2}}
\newcommand{\cthree}{\ion{C}{3}}
\newcommand{\cfour}{\ion{C}{4}}
\newcommand{\none}{\ion{N}{1}}
\newcommand{\ntwo}{\ion{N}{2}}
\newcommand{\nthree}{\ion{N}{3}}

\newcommand{\nfive}{\ion{N}{5}}
\newcommand{\oone}{\ion{O}{1}}
\newcommand{\otwo}{\ion{O}{2}}

\newcommand{\mgtwo}{\ion{Mg}{2}}

\newcommand{\altwo}{\ion{Al}{2}}

\newcommand{\sitwo}{\ion{Si}{2}}
\newcommand{\sithree}{\ion{Si}{3}}
\newcommand{\sifour}{\ion{Si}{4}}

\newcommand{\stwo}{\ion{S}{2}}
\newcommand{\sthree}{\ion{S}{3}}
\newcommand{\sfour}{\ion{S}{4}}

\newcommand{\fetwo}{\ion{Fe}{2}}
\newcommand{\fethree}{\ion{Fe}{3}}

\newcommand{\ebv}{\emph{E(\bv)}\/}

\newcommand{\kms}{km s$^{-1}$}
\newcommand{\logg}{$\log g$}

\newcommand{\msun}{$M_{\sun}$}

\newcommand{\teff}{$T_{\rm eff}$}

\newcommand{\fuse}{{\em FUSE}}
\newcommand{\hst}{{\em HST}}

\shorttitle{\aastex\ Barnard 29 in M13}
\shortauthors{Dixon et al.}

\begin{document}

\title{Observations of the Ultraviolet-Bright Star Barnard 29 in the Globular Cluster M13 (NGC~6205)}

\correspondingauthor{William V. Dixon}
\email{dixon@stsci.edu}

\author[0000-0001-9184-4716]{William V. Dixon}
\affil{Space Telescope Science Institute, 3700 San Martin Drive, Baltimore, MD 21218, USA}

\author[0000-0001-7653-0882]{Pierre Chayer}
\affil{Space Telescope Science Institute, 3700 San Martin Drive, Baltimore, MD 21218, USA}

\author[0000-0003-0531-8547]{I.~N.~Reid}
\affil{Space Telescope Science Institute, 3700 San Martin Drive, Baltimore, MD 21218, USA}

%
\author[0000-0001-8031-1957]{Marcelo Miguel Miller Bertolami}
\affiliation{Instituto de Astrof\'{i}sica de La Plata, UNLP-CONICET, Paseo del Bosque s/n, 1900 La Plata, Argentina}
%




\begin{abstract}

We have analyzed {\em FUSE}, COS, GHRS, and Keck HIRES spectra of the UV-bright star Barnard~29 in M13 (NGC 6205).  By comparing the photospheric abundances derived from multiple ionization states of C, N, O, Si, and S, we infer an effective temperature \teff\ = $21,400 \pm 400$ K.  Balmer-line fits yield a surface gravity \logg\ = $3.10 \pm 0.03$.  We derive photospheric abundances of He, C, N, O, Mg, Al, Si, P, S, Cl, Ar, Ti, Cr, Fe, Ni, and Ge.  Barnard~29 exhibits an abundance pattern typical of the first-generation (FG) stars in M13, enhanced in oxygen and depleted in aluminum.  An underabundance of C and an overabundance of N suggest that the star experienced nonconvective mixing on the RGB.  We see no evidence of significant chemical evolution since the star left the RGB; in particular, it did not undergo third dredge-up.  Previous workers found that the star's FUV spectra yield an iron abundance about 0.5 dex lower than its optical spectrum, but the iron abundances derived from all of our spectra are consistent with the cluster value.  We attribute this difference to our use of model atmospheres without microturbulence, which is ruled out by careful fits to optical absorption features.  We derive a mass $M_*/M_{\sun} = 0.45 - 0.55$ and luminosity $\log L_*/L_{\sun} = 3.26 - 3.35$.  Comparison with stellar-evolution models suggests that Barnard~29 evolved from a ZAHB star of mass $M_*/M_{\sun}$ between 0.50 and 0.55, near the boundary between the extreme and blue horizontal branches.

\end{abstract}





\section{Introduction} \label{sec_intro}

The bright, blue star Barnard~29 in Messier 13 (NGC 6205) has intrigued astronomers for more than a century.  In a series of papers, \citet{Barnard:1900, Barnard:1909, Barnard:1914} pointed out that some of the cluster's stars ``shine with a much bluer light than the great majority'' and cited Barnard~29 as the ``most striking example'' of this group.  \citet{Smith:2005} reviews the early research into these hot, luminous objects, which are now referred to as UV-bright stars.  Some are post-AGB stars, evolving from the asymptotic giant branch (AGB) to the white-dwarf cooling curve at high luminosity; others are AGB-manqu\'{e} stars, evolving directly from the extreme horizontal branch (EHB) at lower luminosity.  Modern spectroscopic studies of Barnard~29 include those of \citet{Conlon:94}, \citet{Dixon:Hurwitz:98}, \citet{Moehler:98}, and \citet{Thompson:07}.  They established the star as a post-AGB object with \teff\ = $20,000 \pm 1000$ K, \logg\ = $2.95 \pm 0.1$, and \abund{He} = $-1.06 \pm 0.20$ and determined its photospheric abundances of C, N, O, Mg, Al, Si, S, and Fe.  Its radial velocity ($V_{\rm LSR} = -228$ \kms; \citealt{Thompson:07}) is consistent with the cluster mean ($V_{\rm LSR} = -228.0$ \kms; \citealt{Harris:96, Harris:2010}).

Despite over a century of study, important questions about Barnard~29 remain unanswered.  In particular, the iron abundance derived from the star's far-ultraviolet (FUV) spectrum is roughly 0.5 dex lower than that derived from optical data, which is consistent with the cluster mean \citep{Thompson:07}.  Another question concerns the extent to which the star's photospheric abundances have been altered by evolutionary processes, particularly nucleosynthesis on the AGB.  The availability of high-resolution, high signal-to-noise FUV and optical spectra allow us to confirm the star's iron abundance and determine the abundances of many additional species.  We employ archival spectra from the {\em Far Ultraviolet Spectroscopic Explorer (FUSE)}, the Cosmic Origins Spectrograph (COS) and Goddard High Resolution Spectrometer (GHRS) aboard the {\em Hubble Space Telescope}, and the Keck High Resolution Echelle Spectrometer (HIRES).  A summary of these data is presented in Table~\ref{tab:log_obs}.  {A link to the \fuse\/ and \hst\/ data is provided here: \dataset[10.17909/T9ZH6R]{http://doi.org/10.17909/T9ZH6R}.}

In Section \ref{sec_observations}, we describe these observations and our reduction of the data.  We present our determination of the star's effective temperature and chemical abundances in Section \ref{sec_teff} and its surface gravity and helium abundance in Section \ref{sec_gravity}.  We describe our model atmospheres in Section \ref{sec_models} and our fitting procedure in Section \ref{sec_fitting}.  We discuss challenges presented by specific stellar features in Section \ref{sec_models_discussion} and examine the implications of our photospheric abundances in Section \ref{sec_abundance_discussion}.  We estimate the star's mass, radius, and luminosity in Section \ref{sec_mass} and discuss  its evolutionary status in Section \ref{sec_evolution}.  Finally, we summarize our conclusions in Section \ref{sec_conclusions}.

\begin{deluxetable*}{lcccclcc}
\tablecaption{Summary of {\it FUSE}, COS, GHRS, and Keck Observations \label{tab:log_obs}}
\tablehead{
\colhead{Instrument} & \colhead{Grating} & \colhead{Wavelength} & \colhead{$R\equiv\lambda/\Delta\lambda$} & \colhead{Exp. Time} & \colhead{Obs.\ Date} & \colhead{Data ID} & \colhead{P.I.} \\
& & \colhead{(\AA)} & & \colhead{(s)}
}
\startdata
{\it FUSE} & $\cdots$ &\phn905--1187 & 20,000 &   15,572 & 2000 Aug 3 & P1015201 & Sembach \\
COS & G130M & 1130--1430 & 10,000 & \phn\phn\phd550 & 2010 Jul 14 & LB2401020 & Green \\
COS & G160M & 1410--1780 & 18,000 & \phn\phn\phd625 & 2010 Jul 14 & LB2401010 & Green \\
GHRS & G200M & 1865--1905 & 20,000 & \phn\phd4570 & 1996 Nov 30 & Z3EC0204T & Napiwotzki \\
HIRES & $\cdots$ & 4300--6700 & 47,800 & \phn\phd1500 & 1996 Jun 5 & HI.19960605.40896 & Reid \\
\enddata
\end{deluxetable*}

\section{Observations and Data Reduction}\label{sec_observations}

\subsection{{\em FUSE}\/ Spectroscopy}

The {\em Far Ultraviolet Spectroscopic Explorer (FUSE)}\/ provides medium-resolution spectroscopy from 1187 \AA\ to the Lyman limit \citep{Moos:00, Sahnow:00}.  Barnard 29 was observed through the \fuse\/ $30\arcsec \times 30\arcsec$ aperture.  The data were reduced using v3.2.1 of CalFUSE, the standard data-reduction pipeline software \citep{Dixon:07}.  CalFUSE corrects for a variety of instrumental effects, extracts spectra from each of the four \fuse\/ channels, and performs wavelength and flux calibration.  The extracted spectra are binned by 0.013 \AA, which corresponds to about 2 detector pixels, or one-fourth of a point-source resolution element.  The spectra from each exposure are aligned by cross-correlating on the positions of stellar absorption features and combined into a single spectrum for each channel.  The signal-to-noise ratio (S/N) per resolution element is about 20 at wavelengths shorter than 1000 \AA\ and about 30 at longer wavelengths.  The spectrum shows absorption from C, N, O, Al, Si, P, S, Cl, Ar, Cr, Fe, Ni, and Ge.  

\citet{Lehner:2004} used this spectrum in a survey designed to study the rate of radiative cooling in the diffuse interstellar medium (ISM).  They combined column-density measurements of \ctwo *,  \stwo, \ion{P}{2}, and \fetwo\ with \hone\ 21 cm emission measurements to derive the cooling rates and analyze the ionization structure, depletion, and metallicity  of  low-, intermediate-, and high-velocity clouds (LVCs, IVCs, and HVCs) along a number of high-latitude sight lines.

\subsection{COS Spectroscopy}

The Cosmic Origins Spectrograph (COS) enables high-sensitivity, medium- and low-resolution spectroscopy in the 1150--3200 \AA\ wavelength range \citep{Green:COS:2012}.  Barnard~29 was observed with COS using both the G130M and G160M gratings. The fully-reduced spectra, processed with CALCOS version 3.2.1 \citep{Fox:2015}, were retrieved from the Mikulski Archive for Space Telescopes (MAST).  The S/N varies between 45 and 65 per 7-pixel resolution element in the G130M spectrum.  In the G160M spectrum, the S/N falls monotonically from $\sim 60$ at 1430 \AA\ to $\sim 20$ at 1770 \AA.  The spectrum shows absorption from He, C, N, O, Mg, Al, Si, S, Ti, Cr, Fe, Ni, and Ge.  

The G160M exposures were obtained first, in time-tag mode, followed by a pair of G130M exposures.  Because the star is bright at short wavelengths, the G130M exposures were obtained in ACCUM mode, which can accommodate higher count rates.  The use of ACCUM mode had two effects on the data:  First, the resolution of the G130M spectrum is significantly lower than the nominal value for this grating ($R \sim 18,000$); modeling the line-spread function with a Gaussian yields FWHM = 0.13 \AA\ ($R \sim 10,000$).  The motion of the COS Optics Select Mechanism 1 (OSM1), which is used to switch between gratings, is not perfectly controlled and can continue throughout an exposure, which in turn moves the spectrum across the detector in an effect known as OSM drift \citep{White:2016}.  Time-tag data can be corrected for OSM drift (CALCOS does this automatically), but ACCUM data cannot, resulting in a lower spectral resolution for the G130M spectrum.  

The second effect is an error in the wavelength solution for the first G130M exposure.  The spectra produced by the two exposures are nearly identical on detector segment B (1130 -- 1280 \AA), but are offset by about 6 pixels at the short-wavelength end of segment A (1290 -- 1430 \AA) and by more than 40 pixels -- nearly 0.5 \AA\ -- at the long-wavelength end.  When taking data in ACCUM mode, COS follows each external exposure with a short exposure of a wavelength-calibration lamp; perhaps the first cal-lamp spectrum was distorted by OSM drift.  We correct for the wavelength error by determining the shift between the two spectra in 10 \AA\ bins across segment A, fitting a linear function to the measured offsets, and rescaling the wavelength array of the first exposure.  Rather than combining the two spectra, we feed both of them to our fitting routines and fit them simultaneously.

\citet{Welsh:2011} used the COS and \fuse\/ spectra to study the ISM toward M13.  They detected absorption due to \cone, \ctwo, \ctwo *, \cfour, \none, \ntwo, \nfive, \oone, \ion{Al}{2}, \sitwo, \sifour, \stwo, and \fetwo\ at $V_{\rm LSR} \sim -60$ \kms, associated with a well-known IVC, and absorption due to \ctwo, \cthree, \cfour, \ntwo, \sitwo, and \sifour\ at $V_{\rm LSR} = -121 \pm 3$ \kms, which they identified as a previously unknown highly-ionized, multi-phase HVC.

\begin{figure}
\plotone{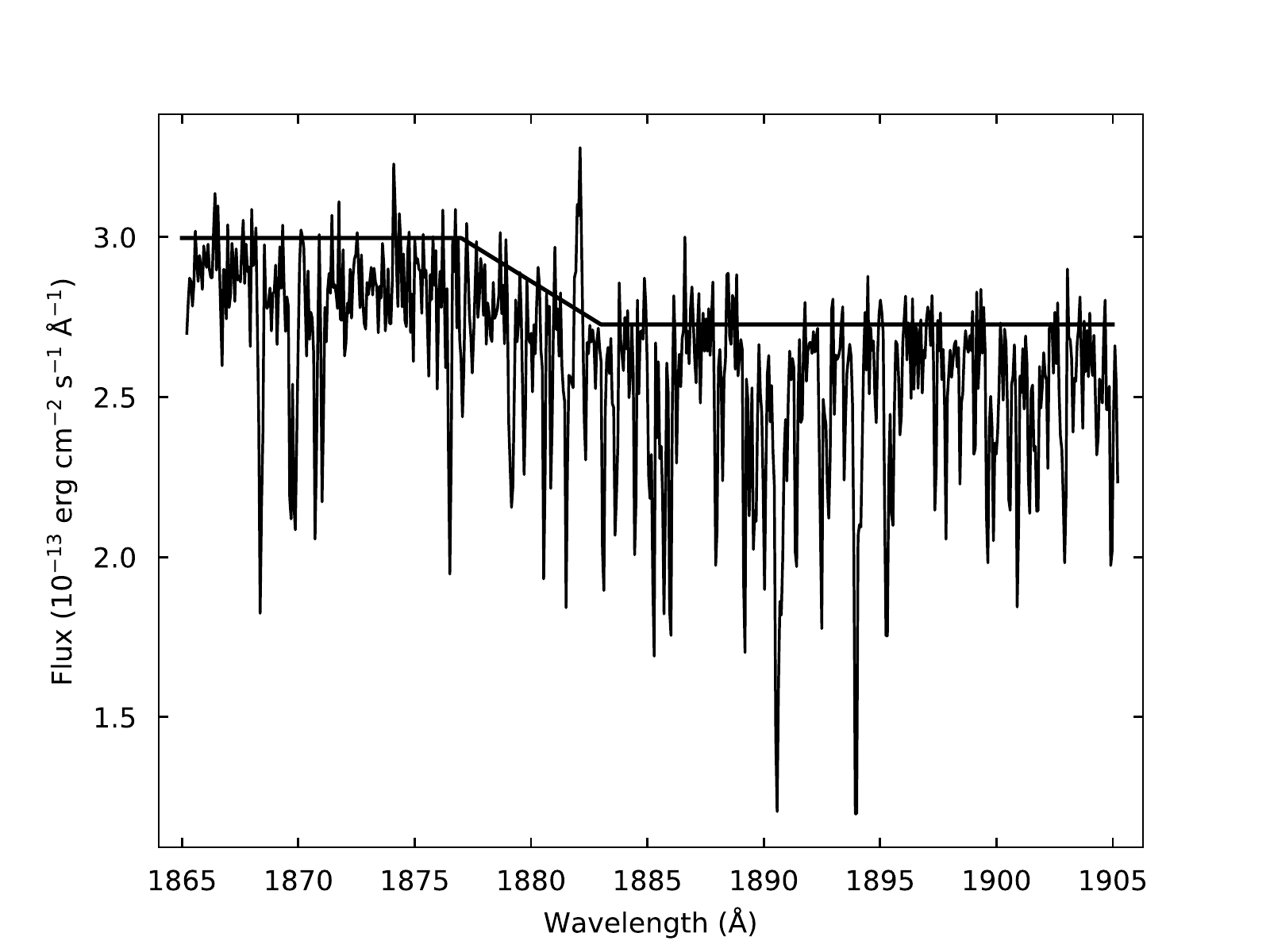}
\caption{GHRS spectrum of Barnard 29, overplotted with our estimate of the continuum.  For this figure, the data are binned by three pixels.  Wavelengths are in the observed frame.}
\label{fig_ghrs}
\end{figure}

\subsection{GHRS Spectroscopy}

The Goddard High-Resolution Spectrograph (GHRS) was one of the four original axial instruments aboard the {\em Hubble Space Telescope}\/ \citep{Heap:95}.  Barnard~29 was observed through the Large Science Aperture (LSA) with the G200M grating.  The fully-reduced spectrum, a product of the final re-calibration of the GHRS data set, was retrieved from MAST.  The spectrum, reproduced in \figref{fig_ghrs}, spans the wavelength range 1865--1905 \AA, a bandpass chosen to enable a measurement of the star's iron abundance \citep{Moehler:98}.  According to version 6.0 of the GHRS Instrument Handbook \citep{GHRSv6}, the resolution of the G200M grating is $R = 20,000$ at 1900 \AA, so we model the line-spread function with a Gaussian of FHWM = 0.095 \AA.  The S/N is about 30 per resolution element across the bandpass.  

The GHRS spectrum of Barnard~29 shows considerable structure, with a jump in the continuum level between about 1877 and 1885 \AA\ (\figref{fig_ghrs}).  The spectrum has been analyzed by both \citet{Moehler:98} and \citet{Thompson:07}, who treat the jump in different ways.  \citeauthor{Moehler:98} assume that it is instrumental.  They define the continuum ``by eye'' and remove it in their normalization of the spectrum.  \citeauthor{Thompson:07} assume that the jump is astrophysical and employ models with a sloping continuum.  Fig.\ 3 of \citeauthor{Thompson:07} shows that the jump is present in the GHRS spectra of both Barnard~29 and ROA~5701, which were observed using the same instrumental set up.  We conclude that the jump is instrumental and follow \citeauthor{Moehler:98} in removing it.  Our estimate of the continuum is overplotted in \figref{fig_ghrs}.

\subsection{Keck Spectroscopy}

Barnard~29 was observed using the HIRES echelle spectrograph on the Keck~I telescope \citep{Vogt:1994}.  The spectrograph was configured to use the red cross-disperser and a slit of width 0\farcs861 and length 7\farcs0. We retrieved the resulting spectrum from the Keck Observatory Archive (KOA). The standard KOA extraction provides one-dimensional spectra that are flat-fielded, bias and background subtracted, and wavelength calibrated.  The spectrum ranges from 4288 to 6630 \AA\ and is divided into 30 spectral orders that cover 70 \AA\ on average. At wavelengths longer than 5200 \AA, there are gaps between the spectral orders that increase in size from a few \AA ngstroms to about 25 \AA\ as the wavelength increases.  The S/N within a 0.1 \AA\ resolution element varies from about 100 to 150 across each spectral order.  The \ion{He}{1} $\lambda$5875 line is not used in the spectral analysis, because it lies at the edge of spectral order 23.  Besides \hone\ and \heone, the Keck spectrum exhibits features of \ntwo, \otwo, \mgtwo, \ion{Al}{3}, \sitwo, \ion{Si}{3}, \ion{S}{2}, \ion{Ar}{2}, and \fethree.  The full list of spectral features and their equivalent widths is presented in Table \ref{tab:lines_hires}. 

The orders of an echelle spectrum have a distinctive shape, which we must remove before attempting to fit individual features.
For each order, we model the continuum using the interactive Python tool specnorm.py. 
The  user identifies spectral regions free of absorption features to which the program fits a third-order spline function.  
The technique does not work for the orders that contain hydrogen Balmer lines, as they are so deep and wide that it is impossible to reconstruct the underlying continuum.
We therefore modify the routine, enabling it to read, write, and manipulate the spline functions.  
For H$\alpha$ and H$\beta$, we instruct the program to read the spline fits to the preceding and subsequent orders,
average them, and scale the result to match a region far from the hydrogen line center.
For H$\alpha$, which falls in order 30, we define an alternative continuum by rescaling the spline fit to order 28.
For H$\beta$, which falls in order 11, we derive an alternative continuum from order 13.
Because order 1 is incomplete, we cannot use it to model the continuum for the H$\gamma$ line, which falls in order 2.
Instead, we use a scaled fit to order 3 as the continuum model and to order 4 as the alternative.  
We create two normalized spectra: for the first, we divide by the continuum model; for the second, we divide by the alternative continuum.  We use the alternative normalization to estimate the uncertainties in our stellar parameters due to errors in our continuum fits.

\startlongtable
\begin{deluxetable}{llrrr}
\tablecaption{Photospheric Lines in the HIRES Spectrum of Barnard~29 \label{tab:lines_hires}}
\tablehead{
\colhead{Ion} & \colhead{$\lambda_{\rm lab}$} & \colhead{$\log gf$} & \colhead{$E_l$} & {EW (m\AA)} \\
\colhead{} & \colhead{(\AA)} & \colhead{} & \colhead{(cm$^{-1}$)} 
}
\startdata
\ion{H}{1} & 4340.462 & $-$0.447 & 82259.102 & $ 2574.9 \pm 43.8 $ \\
              & 4861.323 & $-$0.020 & 82259.102 & $ 2818.8 \pm 70.4 $ \\
              & 6561.010 & 0.710 & 82281.664 & $ 1930.1 \pm 30.4 $ \\
\ion{He}{1}   & 4387.929 & $-$0.883 & 171135.000 & $ 521.3 \pm 8.2 $ \\
              & 4437.551 & $-$2.034 & 171135.000 & $ 102.3 \pm 2.1 $ \\
              & 4471.469 & $-$2.198 & 169086.859 & $ 742.8 \pm 11.1 $ \\
              & 4471.473 & $-$0.278 & 169086.859  \\
              & 4471.473 & $-$1.028 & 169086.859  \\
              & 4471.485 & $-$1.028 & 169086.938  \\
              & 4471.488 & $-$0.548 & 169086.938  \\
              & 4713.139 & $-$1.233 & 169086.859 & $ 222.1 \pm 2.4 $ \\
              & 4713.156 & $-$1.453 & 169086.938  \\
              & 4713.376 & $-$1.923 & 169087.922  \\
              & 4921.931 & $-$0.435 & 171135.000 & $ 505.1 \pm 6.2 $ \\
              & 5015.678 & $-$0.820 & 166277.547 & $ 217.5 \pm 2.3 $ \\
              & 5047.738 & $-$1.602 & 171135.000 & $ 151.0 \pm 2.8 $ \\
              & 5875.599\tablenotemark{a} & $-$1.511 & 169086.859  \\
              & 5875.614\tablenotemark{a} & $-$0.341 & 169086.859  \\
              & 5875.615\tablenotemark{a} & 0.409 & 169086.859  \\
              & 5875.625\tablenotemark{a} & $-$0.341 & 169086.938  \\
              & 5875.640\tablenotemark{a} & 0.139 & 169086.938  \\
              & 5875.966\tablenotemark{a} & $-$0.211 & 169087.922  \\
              & 6678.154\tablenotemark{b} & 0.329 & 171135.000 & $ 474.5 \pm 4.2 $ \\
\ion{N}{2}    & 4432.736 & 0.580 & 188857.375 & $ 12.0 \pm 1.4 $ \\
              & 4447.030 & 0.228 & 164610.760 & $ 32.1 \pm 1.4 $ \\
              & 4530.410 & 0.670 & 189335.156 & $ 12.2 \pm 1.1 $ \\
              & 4552.522\tablenotemark{b} & 0.329 & 189335.156 \\
              & 4601.478 & $-$0.428 & 148940.170 & $ 27.0 \pm 1.8 $ \\
              & 4607.153 & $-$0.507 & 148908.590 & $ 26.3 \pm 1.4 $ \\
              & 4613.868 & $-$0.665 & 148940.170 & $ 17.9 \pm 1.5 $ \\
              & 4621.393 & $-$0.514 & 148940.170 & $ 17.2 \pm 1.2 $ \\
              & 4630.539 & 0.094 & 149076.520 & $ 48.8 \pm 1.5 $ \\
              & 4643.086 & $-$0.359 & 149076.520 & $ 29.1 \pm 1.3 $ \\
              & 4788.138 & $-$0.363 & 166582.450 & $ 9.8 \pm 1.4 $ \\
              & 4803.287 & $-$0.113 & 166678.640 & $ 16.8 \pm 1.3 $ \\
              & 4994.360 & $-$0.164 & 205654.220 & $ 13.3 \pm 1.3 $ \\
              & 4994.370 & $-$0.069 & 168892.210 \\
              & 5001.134 & 0.263 & 166521.690 & $ 29.3 \pm 1.3 $ \\
              & 5001.474 & 0.441 & 166582.450 & $ 36.5 \pm 1.2 $ \\
              & 5002.703 & $-$1.022 & 148908.590 & $ 11.1 \pm 1.1 $ \\
              & 5005.150 & 0.594 & 166678.640 & $ 46.3 \pm 1.2 $ \\
              & 5007.328 & 0.171 & 168892.210 & $ 17.1 \pm 1.0 $ \\
              & 5010.621 & $-$0.607 & 148940.170 & $ 19.4 \pm 1.1 $ \\
              & 5016.381 & $-$0.515 & 166582.450 & $ 10.8 \pm 1.2 $ \\
              & 5025.659 & $-$0.547 & 166678.640 & $ 5.3 \pm 0.9 $ \\
              & 5045.099 & $-$0.407 & 149076.520 & $ 28.7 \pm 1.9 $ \\
              & 5495.655 & $-$0.265 & 170666.230 & $ 9.6 \pm 1.1 $ \\
              & 5666.630 & $-$0.045 & 148940.170 & $ 37.6 \pm 1.3 $ \\
              & 5676.020 & $-$0.367 & 148908.590 & $ 22.8 \pm 1.2 $ \\
              & 5679.560 & 0.250 & 149076.520 & $ 56.3 \pm 1.3 $ \\
              & 5686.210 & $-$0.549 & 148940.170 & $ 11.8 \pm 1.0 $ \\
              & 5710.770 & $-$0.518 & 149076.520 & $ 21.6 \pm 1.7 $ \\
              & 5747.300 & $-$1.075 & 149187.800 & $ 10.0 \pm 1.0 $ \\
              & 5931.780\tablenotemark{b} & 0.052 & 170607.890 & $ 24.2 \pm 1.8 $ \\
              & 5940.240 & $-$0.445 & 170607.890 & $ 11.1 \pm 1.3 $ \\
              & 5941.650 & 0.313 & 170666.230 & $ 22.0 \pm 1.3 $ \\
              & 6379.620 & $-$0.951 & 148940.170 & $ 8.2 \pm 1.2 $ \\
              & 6482.050 & $-$0.245 & 149187.800 & $ 24.6 \pm 1.6 $ \\
\ion{O}{2}    & 4317.139 & $-$0.386 & 185235.281 & $ 9.4 \pm 1.4 $ \\
              & 4319.630 & $-$0.380 & 185340.577 & $ 12.0 \pm 1.3 $ \\
              & 4345.560 & $-$0.346 & 185340.577 & $ 7.3 \pm 1.2 $ \\
              & 4349.426 & 0.060 & 185499.124 & $ 26.9 \pm 1.3 $ \\
              & 4351.260 & 0.227 & 206971.680 & $ 10.0 \pm 1.1 $ \\
              & 4366.895 & $-$0.348 & 185499.124 & $ 15.1 \pm 1.1 $ \\
              & 4414.899 & 0.172 & 189068.514 & $ 27.7 \pm 1.4 $ \\
              & 4416.975 & $-$0.077 & 188888.543 & $ 16.9 \pm 1.3 $ \\
              & 4443.010 & $-$0.047 & 228723.840 & $ 3.9 \pm 0.9 $ \\
              & 4452.378 & $-$0.788 & 189068.514 & $ 3.8 \pm 1.1 $ \\
              & 4590.974 & 0.350 & 206971.680 & $ 11.7 \pm 1.1 $ \\
              & 4638.856 & $-$0.332 & 185235.281 & $ 15.1 \pm 1.2 $ \\
              & 4641.810 & 0.055 & 185340.577 & $ 27.2 \pm 1.1 $ \\
              & 4649.135 & 0.308 & 185499.124 & $ 38.8 \pm 1.2 $ \\
              & 4650.838 & $-$0.362 & 185235.281 & $ 15.3 \pm 1.2 $ \\
              & 4661.632 & $-$0.278 & 185340.577 & $ 16.0 \pm 1.3 $ \\
              & 4676.235 & $-$0.394 & 185499.124 & $ 13.1 \pm 1.1 $ \\
              & 4705.346 & 0.477 & 211712.732 & $ 9.5 \pm 0.9 $ \\
              & 4924.529 & 0.074 & 212161.881 & $ 5.8 \pm 1.1 $ \\
\ion{Mg}{2}   & 4481.126 & 0.740 & 71490.188 & $ 15.0 \pm 1.0 $ \\
              & 4481.150 & $-$0.560 & 71490.188 \\
              & 4481.325 & 0.590 & 71491.063 & $ 13.4 \pm 1.1 $ \\
\ion{Al}{3}   & 4512.565 & 0.410 & 143633.375 & $ 5.6 \pm 0.9 $ \\
              & 4529.189 & 0.660 & 143713.500 & $ 10.3 \pm 1.0 $ \\
              & 5722.730 & $-$0.070 & 126164.047 & $ 14.2 \pm 1.4 $ \\
%
\ion{Si}{2}   & 5055.984 & 0.593 & 81251.320 & $ -8.2 \pm 1.7 $ \\
              & 6347.109 & 0.297 & 65500.469 & $ -7.4 \pm 1.8 $ \\
              & 6371.371 & $-$0.003 & 65500.469 & $ -3.9 \pm 0.9 $ \\
\ion{Si}{3}   & 4552.622\tablenotemark{b} & 0.181 & 153377.047 & $ 76.1 \pm 1.8 $ \\
              & 4567.840 & $-$0.039 & 153377.047 & $ 57.2 \pm 1.4 $ \\
              & 4574.757 & $-$0.509 & 153377.047 & $ 32.6 \pm 1.4 $ \\
              & 5739.734 & $-$0.160 & 159069.609 & $ 29.6 \pm 1.3 $ \\
\ion{S}{2}    & 4552.410\tablenotemark{b} & $-$0.100 & 121528.719  \\
              & 5639.977 & 0.330 & 113461.539 & $ 7.1 \pm 1.2 $ \\
\ion{Ar}{2}   & 4348.064 & 0.470 & 134241.734 & $ 6.0 \pm 1.2 $ \\
\ion{Fe}{3}   & 4419.596 & $-$2.218 & 66464.641 & $ 5.2 \pm 1.0 $ \\
              & 5156.111 & $-$2.018 & 69695.727 & $ 8.2 \pm 1.3 $ \\
              & 5833.938 & 0.616 & 149285.000 & $ 6.6 \pm 1.4 $ \\
\enddata
\tablenotetext{a}{Line falls on edge of spectral order.} 
\tablenotetext{b}{Feature is blended.}
\end{deluxetable}

\begin{figure}
\plotone{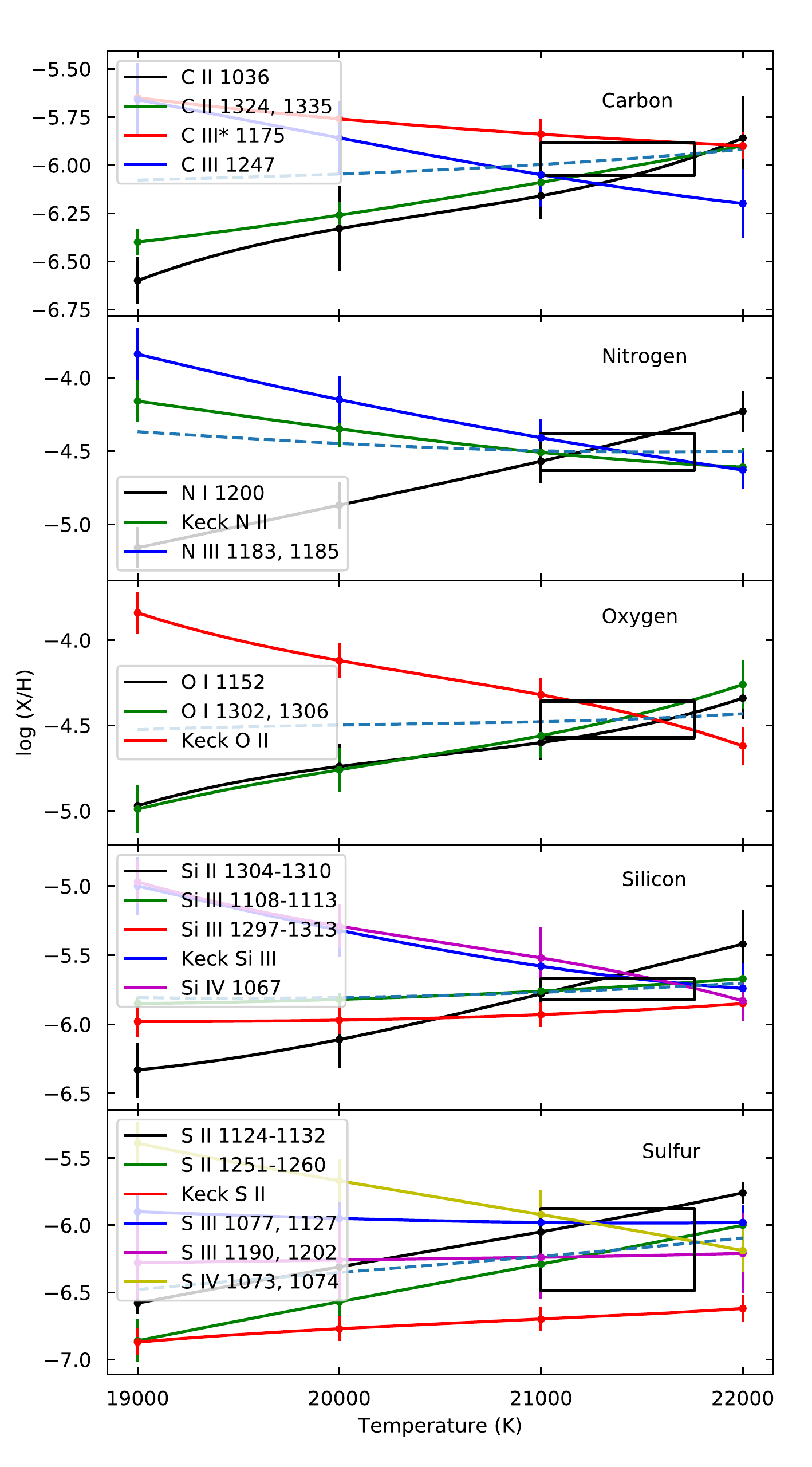}
\caption{Deriving the effective temperature and abundance of C, N, O, Si, and S.  Points with error bars represent the abundance derived from model fits to each absorption feature.  Solid lines are low-order polynomial fits to the measured points.  Dashed lines represent the weighted mean of the solid lines, computed at 10 K intervals.  The black box in each panel denotes the allowed range of temperature and abundance, as described in the text.}
\label{fig_temp_abundance}
\end{figure}

\begin{figure*}
\plotone{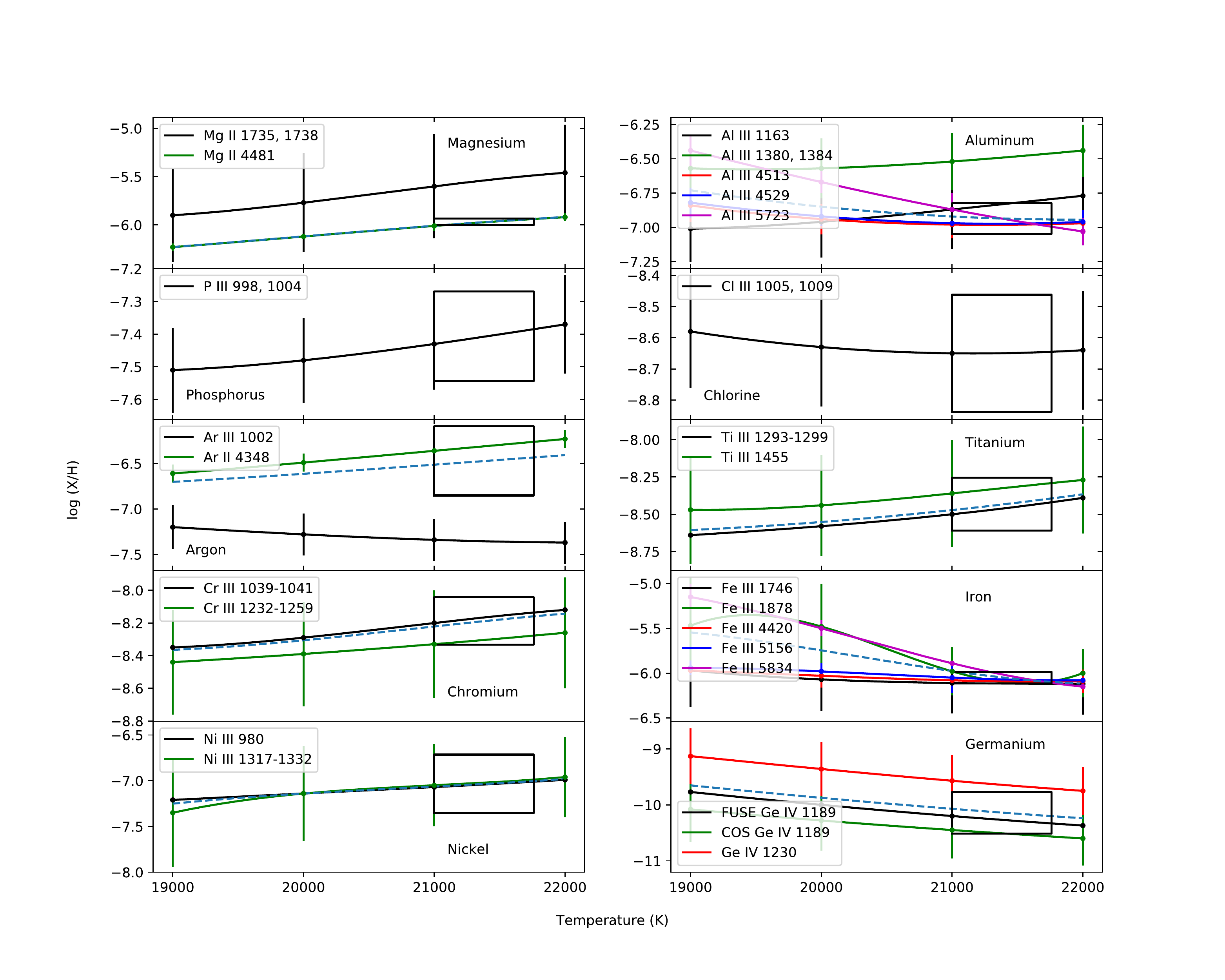}
\caption{Deriving the abundance of eight additional elements.  Points with error bars represent the abundance derived from model fits to each absorption feature.  Solid lines are low-order polynomial fits to the measured points.  Dashed lines represent the weighted mean of the solid lines, computed at 10 K intervals.  The black box in each panel denotes the allowed range of temperature and abundance, as described in the text.}
\label{fig_only_abundance}
\end{figure*}

\begin{deluxetable*}{ll}
\tablecaption{Selected Absorption Features in the FUV Spectrum of Barnard~29 \label{tab:fuv_lines}}
\tablehead{
\colhead{Species} & \colhead{Lines Fit (\AA)}
}
\startdata
Carbon          & \ctwo\ 1036.3, 1324.0, 1334.5; \cthree * 1175 (multiplet); \cthree\ 1247.4  \\
Nitrogen        & \none\ 1199.6; \nthree\ 1183.0, 1184.5 \\
Oxygen          & \oone\ 1152.2, 1302.2, 1306.0 \\
Magnesium	& \ion{Mg}{2} 1734.9, 1737.6 \\
Aluminum    & \ion{Al}{3} 1162.6, 1379.7, 1384.1 \\
Silicon         & \sitwo\ 1304.4, 1309.3, 1309.5 \\
			& \sithree\ 1108.4, 1110.0, 1113.2, 1296.7, 1301.1, 1312.6 \\
			& \sifour\ 1066.6 \\
Phosphorus & \ion{P}{3} 998.0, 1003.6  \\
Sulfur          & \stwo\ 1124.4, 1125.0, 1131.1, 1131.7, 1250.6, 1253.8, 1259.5 \\
		  & \sthree\ 1077.1, 1126.5, 1126.9, 1190.2, 1202.1 \\
                    & \sfour\ 1073.0, 1073.5 \\
Chlorine	& \ion{Cl}{3} 1005.3, 1008.8  \\
Argon	& \ion{Ar}{3} 1002.1  \\
Titanium        &\ion{Ti}{3} 1293.2, 1295.9, 1298.6, 1298.7, 1455.2 \\
Chromium     & \ion{Cr}{3} 1039--1041, 1231.9, 1233.0, 1236.2, 1238.5, 1247.8, 1252.6, 1259.0  \\
Iron            &  \ion{Fe}{3} 1745.6, 1878.0  \\
Nickel          &\ion{Ni}{3} 979.6, 1317--1332 \\
Germanium & \ion{Ge}{4} 1189.0, 1229.8 \\
\enddata
\tablecomments{For chromium and nickel, we fit multiple absorption features in the listed region.}
\end{deluxetable*}

\section{Analysis}\label{sec_analysis}

\subsection{Effective Temperature from Ionization Balance}\label{sec_teff}

We determine the star's effective temperature by comparing the absorption features of multiple ionization states of C, N, O, Si, and S.  
Consider the panel labeled ``Nitrogen'' in \figref{fig_temp_abundance}.  We fit the COS \none\/ $\lambda 1200$ feature with models (described in Section \ref{sec_models}) assuming \teff\ = 19,000 K and find a best-fit value of \abund{N} = $-5.16 \pm 0.14$.  We repeat with models assuming \teff\ = 20,000 K, 21,000 K, and 22,000 K.  The best-fit abundances are plotted as black points and connected with a low-order polynomial (evaluated at 10 K intervals).  Vertical bars represent the statistical uncertainties returned by our fitting routine.  As the temperature rises, the fraction of nitrogen in the neutral state falls, requiring a higher nitrogen abundance to reproduce the observed feature.  Our Keck spectrum exhibits a number of \ntwo\ lines (Table \ref{tab:lines_hires}).  We fit each of them individually and plot the mean abundance as a function of temperature in green.  We fit the \nthree\ $\lambda \lambda 1183, 1185$ doublet in the \fuse\/ spectrum in the same way and plot the results in blue.  (The \nthree\ features also appear in the COS spectrum, but they yield abundances with larger error bars, so we use the \fuse\/ values.)  As the fraction of ionized nitrogen rises with temperature, the best-fit nitrogen abundance falls.

We repeat this process for C, O, Si, and S, fitting all of the features in the Keck spectrum (Table \ref{tab:lines_hires}) and selected features in the FUV (Table \ref{tab:fuv_lines}).  With few exceptions, the abundance curves intersect between 21,000 K and 22,000 K.  To quantify this result, we compute the error-weighted mean abundance and the error-weighted standard deviation as a function of temperature for each element.  The mean abundance is plotted as a dashed line in each panel.  At each temperature step, we compute 
$$\chi^2 = \sum \Bigl \lbrace [y_i - y(x_i)]^2 / \sigma_i^2 \Bigr \rbrace,$$
where $y_i$ is the abundance derived from a single feature (or group of features; solid line), $y(x_i)$ is the mean abundance (dashed line), $\sigma_i$ is the statistical uncertainty in the derived abundance (vertical bars), and the summation is taken over the 21 abundance curves plotted in \figref{fig_temp_abundance}.  $\chi^2$ has a minimum at \teff\ = 21,400 K.  Our lower limit to the temperature is \teff\ = 21,000 K, set by the point at which $\chi^2$ rises by 7.04 relative to its minimum.  This choice of $\Delta \chi^2$ is strictly correct for a model with six interesting parameters (in our case, one temperature and five abundances) if all errors are normally distributed \citep{Press:88}.  For each element, our best-fit abundance is the error-weighted mean value (computed above) at the best-fit temperature.  The abundance uncertainty is the larger of the error-weighted standard deviation or the measured uncertainty for the feature that dominates the fit.  The black box in each panel denotes the allowed range of temperature and abundance for that element.

\begin{figure}
\plotone{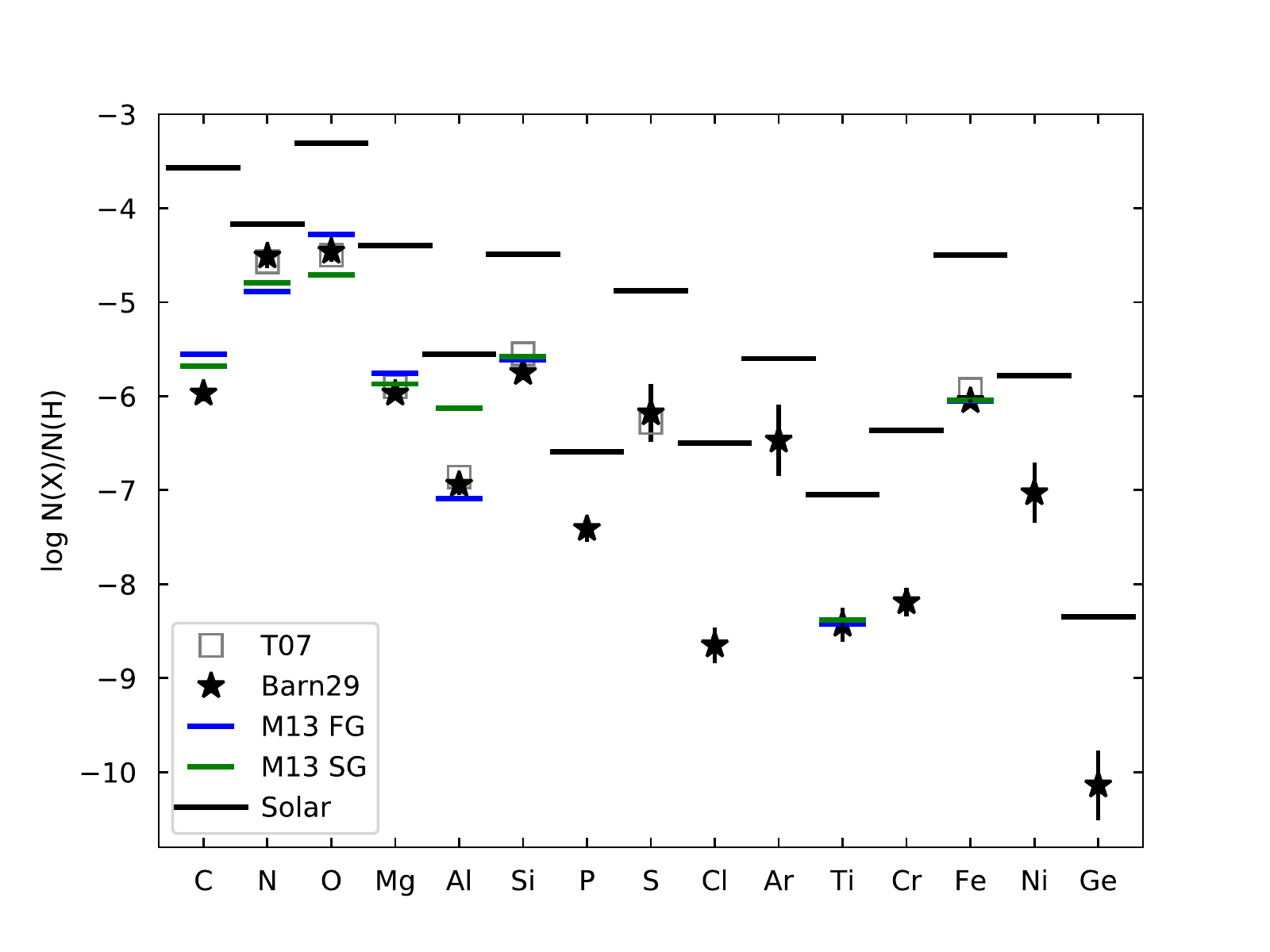}
\caption{Photospheric abundances of Barnard 29 (from Table \ref{tab:table}; stars), the solar photosphere (black lines; \citealt{Asplund:2009}), and the first-generation (FG, blue lines) and second-generation (SG, green lines) stars in M13; \citealt{Meszaros:2015}. Open squares are from fits to the star's optical spectrum by \citealt{Thompson:07}.}
\label{fig_abundance}
\end{figure}

Ten other metals, Mg, Al, P, Cl, Ar, Ti, Cr, Fe, Ni, and Ge, exhibit absorption lines in the spectrum of Barnard~28, but only of a single ionization state.  (Argon has absorption features from \ion{Ar}{2} and \ion{Ar}{3}, but the resulting abundance curves do not cross.)  We compute their abundances just as for the lighter elements, but do not include them in the calculation of the best-fit temperature.  The absorption curves for these species are presented in \figref{fig_only_abundance}.  The full set of abundances for Barnard~29 is presented in Table \ref{tab:table} and plotted in \figref{fig_abundance}.  Our results are generally consistent with those of \citet{Thompson:07}.

\begin{deluxetable*}{lccccccc}
\tablecaption{Photospheric Abundances of Barnard~29, M13, and the Sun \label{tab:table}}
\tablehead{
\colhead{Species} & \colhead{Barnard 29} & \multicolumn{4}{c}{M13} & \colhead{Sun} \\
\cline{3-6}
\colhead{} & \colhead{} & \colhead{FG}  & \colhead{SG} & \colhead{Uncertainty} & \colhead{Scatter}
}
\startdata
Helium & $-0.89 \pm 0.04$ &	\nodata & \nodata & \nodata & \nodata	& $-1.07 \pm 0.01$ \\
Carbon & $-5.97 \pm 0.09$ &	$-5.55$ & $-5.68$ & 0.14 & 0.07	& $-3.57 \pm 0.05$ \\
Nitrogen & $-4.51 \pm 0.13$ &	$-4.89$ & $-4.79$ & 0.13 & 0.17	& $-4.17 \pm 0.05$ \\
Oxygen & $-4.46 \pm 0.11$ &	$-4.28$ & $-4.71$ & 0.13 & 0.17	& $-3.31 \pm 0.05$ \\
Magnesium & $-5.97 \pm 0.04$ &	$-5.76$ & $-5.87$ & 0.06 & 0.15	& $-4.40 \pm 0.04$ \\
Aluminum & $-6.94 \pm 0.11$ &	$-7.09$ & $-6.13$ & 0.13 & 0.53	& $-5.55 \pm 0.03$ \\
Silicon & $-5.75 \pm 0.08$ &	$-5.61$ & $-5.58$ & 0.08 & 0.09	& $-4.49 \pm 0.03$ \\
Phosphorus & $-7.41 \pm 0.14$ &	\nodata & \nodata & \nodata & \nodata	& $-6.59 \pm 0.03$ \\
Sulfur & $-6.18 \pm 0.31$ &	\nodata & \nodata & \nodata & \nodata	& $-4.88 \pm 0.03$ \\
Chlorine & $-8.65 \pm 0.19$ &	\nodata & \nodata & \nodata & \nodata	& $-6.50 \pm 0.30$ \\
Argon & $-6.47 \pm 0.38$ &	\nodata & \nodata & \nodata & \nodata	& $-5.60 \pm 0.13$ \\
Titanium & $-8.43 \pm 0.18$ &	$-8.42$ & $-8.38$ & 0.16 & 0.14	& $-7.05 \pm 0.05$ \\
Chromium & $-8.19 \pm 0.15$ &	\nodata & \nodata & \nodata & \nodata	& $-6.36 \pm 0.04$ \\
Iron & $-6.05 \pm 0.07$ &	$-6.05$ & $-6.04$ & 0.07 & 0.07	& $-4.50 \pm 0.04$ \\
Nickel & $-7.03 \pm 0.32$ &	\nodata & \nodata & \nodata & \nodata	& $-5.78 \pm 0.04$ \\
Germanium & $-10.14 \pm 0.37$ &	\nodata & \nodata & \nodata & \nodata	& $-8.35 \pm 0.10$ \\
\enddata
\tablecomments{Abundances relative to hydrogen: \abund{X}.  M13 values from \citealt{Meszaros:2015}.  FG = first generation; SG = second generation. ``Uncertainty'' represents both random and systematic uncertainties in the derived abundance. ``Scatter'' represents star-to-star scatter within the cluster.  Both value are computed for the entire cluster, rather than either subpopulation.  Solar values from \citealt{Asplund:2009}.}
\end{deluxetable*}

\subsection{Surface Gravity and Helium Abundance}\label{sec_gravity}

\begin{figure}
\plotone{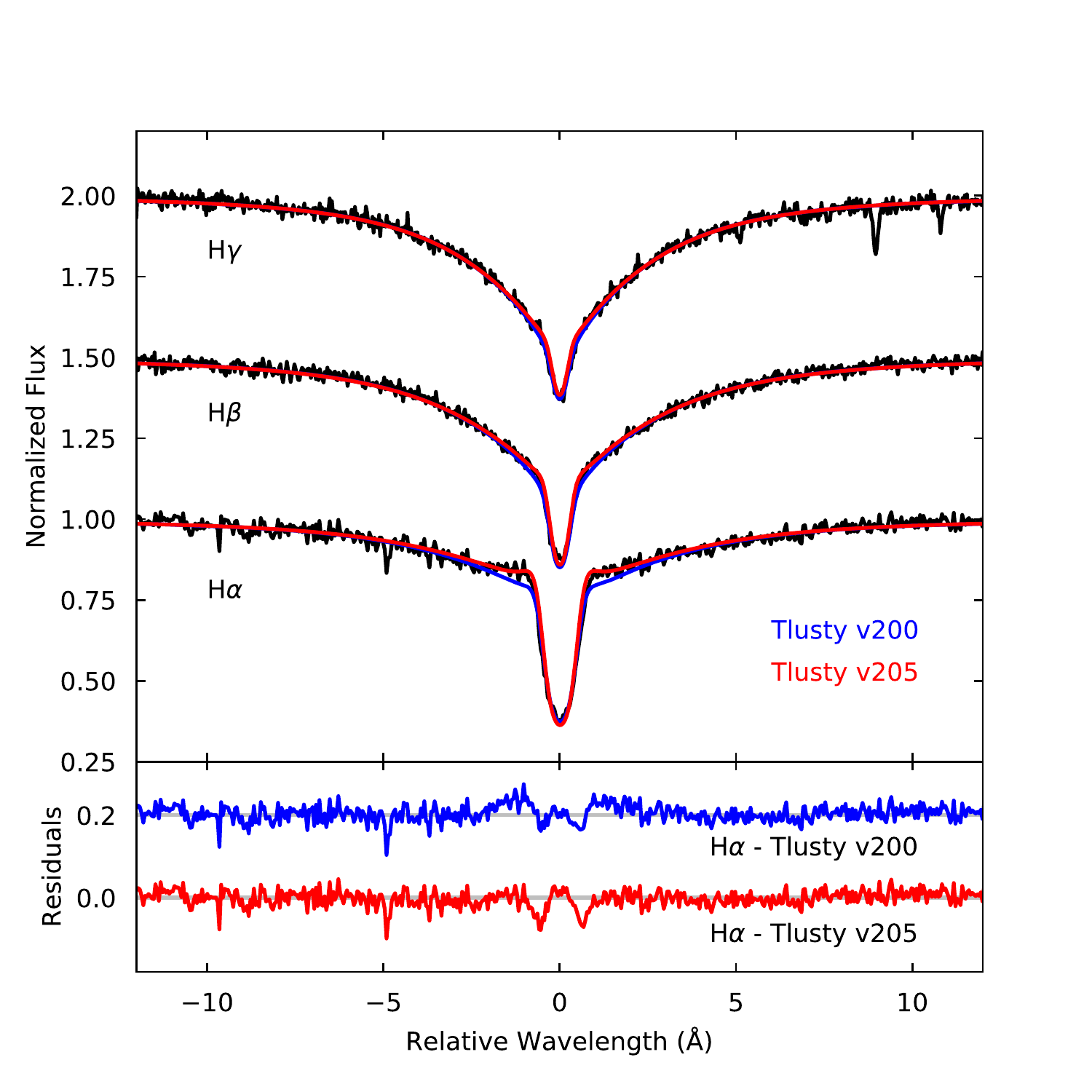}
\caption{Hydrogen lines in the optical spectrum of Barnard 29 (black).  The red curve represents our best-fit synthetic spectrum.  It was derived from a stellar atmosphere computed with TLUSTY v205, which employs hydrogen collisional excitation rates taken from \citealt{Przybilla:Butler:2004}.  The blue curve was derived from an atmosphere generated with TLUSTY v200, which uses the  collisional rates of \citealt{Mihalas:1975}.}
\label{fig_hydrogen}
\end{figure}

\begin{figure}
\plotone{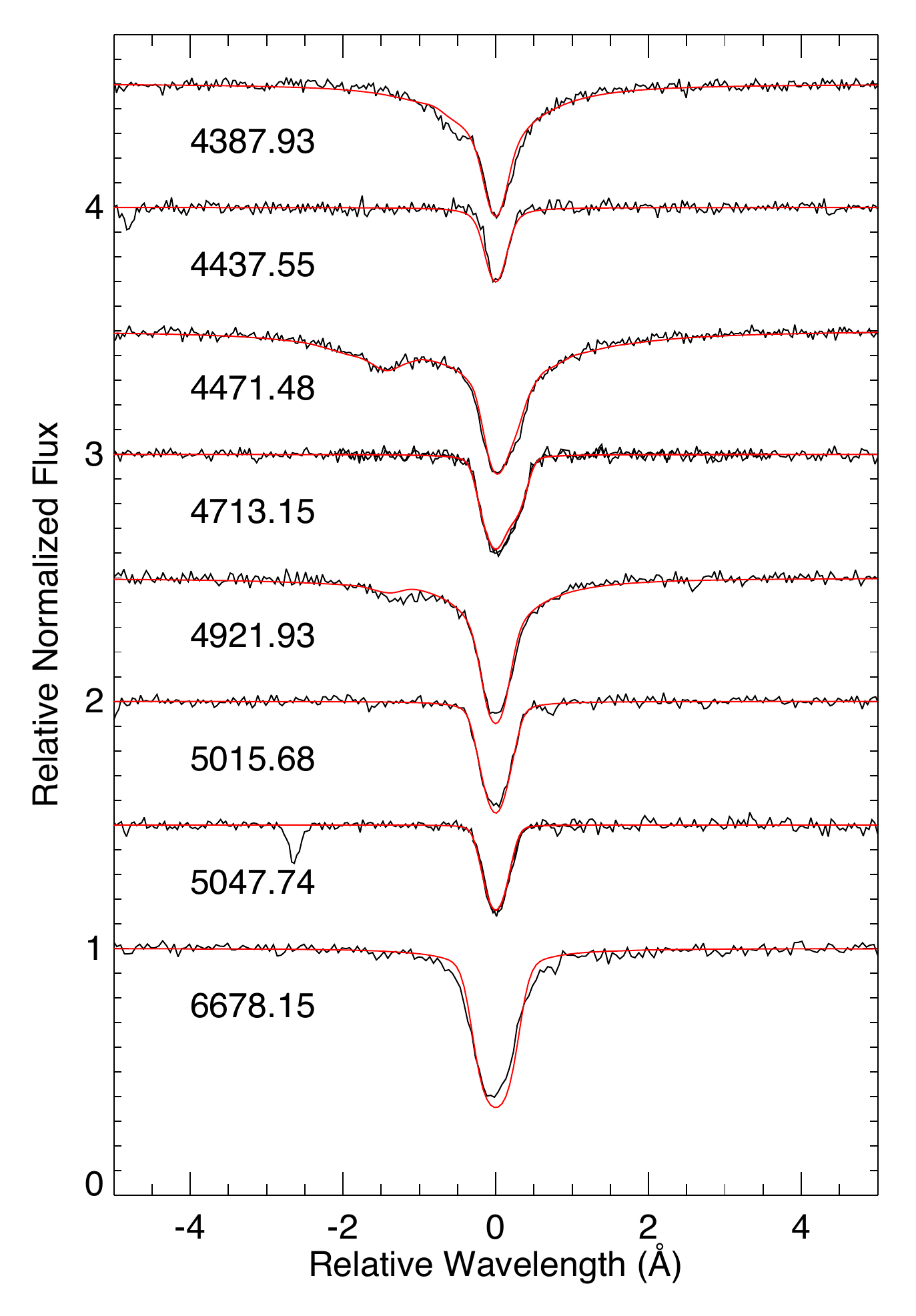}
\caption{\heone\ lines in the optical spectrum of Barnard 29 (black) with our best-fit synthetic spectrum (red).  Two of the lines appear in multiple spectral orders.  The line at 6678 \AA\ appears to blended with another feature and is not used to determine the helium abundance.}
\label{fig_helium}
\end{figure}

We determine the surface gravity from a simultaneous fit to the star's Balmer lines (H$\alpha$, H$\beta$, and H$\gamma$).  Assuming \teff\ = 21,400 K, we find that \logg\ = $3.100 \pm 0.003$.  This value is not sensitive to small errors in the optical continuum:  Repeating the fit using our alternative normalization yields \logg\ = $3.111 \pm 0.003$.  Reducing the temperature to 21,000 K yields \logg\ = $3.071 \pm 0.003$; we will adopt this value as the lower limit to the surface gravity.  We determine the star's helium abundance by individually fitting the seven unblended helium features in Table \ref{tab:lines_hires}.  Models assuming our best-fit effective temperature and surface gravity yield a helium abundance of \abund{He} = $-0.89 \pm 0.04$, where the abundance is the error-weighted mean of our seven results, and the uncertainty is their error-weighted standard deviation.  Reducing both the temperature and gravity to their lower limits does not change this result.  Our final atmospheric parameters are thus \teff\ = $21,400 \pm 400$ K, \logg\ = $3.10 \pm 0.03$, and \abund{He} = $-0.89 \pm 0.04$.  This model successfully reproduces the hydrogen (\figref{fig_hydrogen}) and helium features (\figref{fig_helium}) in our HIRES spectrum. 

Our metal abundances were derived using models with \logg\ = 3.0, but our final value is  \logg\ = 3.1.  How sensitive are our abundances to this change in the surface gravity?  In our fits to the \nthree\ $\lambda \lambda 1183, 1185$ doublet in the \fuse\/ spectrum, we found \abund{N} = $-4.41 \pm 0.16$ using models with \teff\ = 21,000 K and \logg\ = 3.0.  Repeating the fit using models with \logg\ = 3.1 yields \abund{N} = $-4.38 \pm 0.15$.  The difference is smaller than the uncertainty in any of our abundances.

Using \citet{Kurucz:92} LTE models with [Fe/H] $= -1.0$ and scaled-solar abundances, \citet{Conlon:94} derived stellar parameters of \teff\ = $20,000 \pm 1000$ K, \logg\ = $3.0 \pm 0.1$, and \abund{He} = $-1.06 \pm 0.20$ for Barnard~29.  \citet{Thompson:07} used a grid of NLTE models generated with TLUSTY and SYNSPEC, just as we do, but their models included iron with an abundance [Fe/H] $= -1.1$ and the light elements C, N, O, Mg, Si, and S.  They found \teff\ = $20,000 \pm 1000$ K and \logg\ = $2.95 \pm 0.1$.  Both groups determined the star's effective temperature by fitting its \sitwo\ and \sithree\ lines and its surface gravity by fitting its H$\gamma$ line.  These results are generally consistent with ours.  We discuss the use of silicon as temperature indicator in Section \ref{sec_silicon3}.

\subsection{Model Atmospheres}\label{sec_models}

We compute a grid of non-local thermodynamic equilibrium (NLTE) stellar-atmosphere models using version 205 of the program TLUSTY \citep{Hubeny:Lanz:95}.  The models are composed of hydrogen, helium, nitrogen, oxygen and the element whose lines we wish to fit.  The models have effective temperatures \teff\ = 19,000, 20,000, 21,000, and 22,000 K; a surface gravity \logg\ = 3.0; and abundances \abund{He} = $-0.9$, \abund{N} = $-4.3$, and \abund{O} = $-4.5$; these parameters are similar to those derived for Barnard~29 by \citet{Thompson:07}.  When fitting optical hydrogen and helium lines to determine the surface gravity and helium abundance, we use a grid with surface gravities \logg\ = 2.8 to 3.2 in steps of 0.1 dex and helium abundances \abund{He} = $-1.1$ to $-0.7$ in steps of 0.2 dex.  We employ atomic models similar to those used by \citet{Lanz:Hubeny:2007} to compute their grid of B-type stars.  

From these model atmospheres, we compute synthetic spectra using version 51 of the program SYNSPEC \citep{Hubeny:88}.  For the Keck data, the synthetic spectra are convolved with a Gaussian of FWHM = 0.1 \AA\ to replicate the observed spectrum.  For the \fuse\/ data, the synthetic spectra are convolved with a Gaussian of FWHM = 0.06 \AA\ to match the \fuse\/ line-spread function.  For the COS G130M spectrum, we model the line-spread function with a Gaussian of FWHM = 0.13 \AA.  For the G160M spectrum, we employ the tabulated line-spread functions appropriate for data obtained at Lifetime Position \#1, which are available from \href{http://www.stsci.edu/hst/cos/performance/spectral_resolution/}{the COS website}.  For the GHRS spectrum, we adopt a Gaussian with FWHM = 0.095 \AA\ as the line-spread function.

For the elements P, Cl, Ti, Cr, Ni, and Ge, we lack the model atoms necessary to compute full NLTE models.  For these species, we adopt an NLTE model atmosphere with a H+He+N+O+Fe composition and compute the ionization fractions of the element in question assuming LTE.  We assume \abund{Fe} = $-6.0$, which is roughly the cluster abundance.  For the iron-peak elements Ti, Cr, and Ni, we can use iron, for which we have a complete set of models, to estimate the abundance errors.  To this end, we generate a second grid of iron spectra using the LTE approximation.  According to TLUSTY, the dominant ionization state of iron in the stellar photosphere is \ion{Fe}{3}.  Fitting our LTE models to the star's \ion{Fe}{3} lines yields an abundance consistent with that derived from the NLTE models.  \ion{Ti}{3}, \ion{Cr}{3}, \ion{Fe}{3} and \ion{Ni}{3} have ionization energies of 27, 31, 31, and 35 eV, respectively, so we expect their populations to vary similarly with temperature.

SYNSPEC computes partition functions for only the first three ionization states of elements heavier than nickel.  As a result, there are by default no \ion{Ge}{4} features in our synthetic spectra.  To derive an abundance from the star's \ion{Ge}{4} lines, we must modify the program to include the ionization potentials and partition functions of \ion{Ge}{4} and \ion{Ge}{5}.  We compute the partition functions by considering all of the statistical weights and energy levels provided by \citet{Sugar:Musgrove:1993}. The atomic data for the \ion{Ge}{4} 1189 \AA\ and 1230 \AA\ transitions are from \citet{Morton:2000}.

\subsection{Spectral Fitting}\label{sec_fitting}

Models are fit to the data using a chi-squared minimization routine.  Because both the data and the synthetic spectra are normalized, the only free parameter is the abundance of the element in question.  Our program linearly interpolates between two model spectra to compute a model with an intermediate abundance.  The abundance uncertainties quoted for individual line fits are one-standard deviation errors computed from the covariance matrix returned by the fitting routine; we refer to these as statistical errors.   

\begin{figure}
\plotone{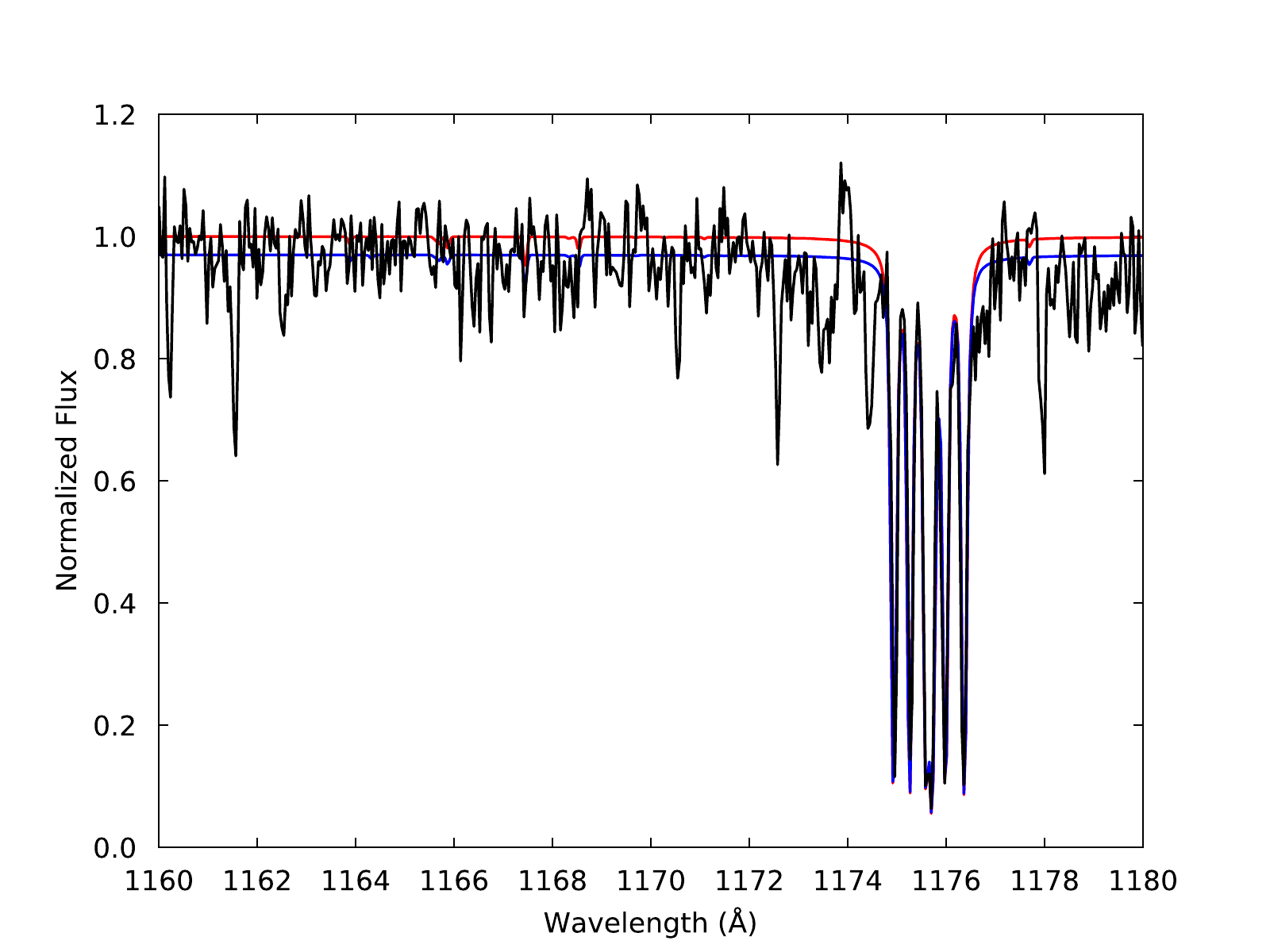}
\caption{\cthree * lines in the \fuse\/ spectrum of Barnard 29.   The data are normalized as described in the text.  The red curve represents a model whose continuum is fixed at 1.0.  The blue curve represents a model whose continuum is fixed at 0.97.  For this figure, the spectrum has been normalized, binned by three pixels, and shifted to a laboratory wavelength scale.}
\label{fig_carbon3}
\end{figure}

Continuum placement is the dominant uncertainty in our fits to the \fuse\/ and COS spectra.  The FUV spectra of early B-type stars are riddled with absorption lines, and even relatively flat regions of the spectrum are depressed by a myriad of weak absorption features.  Allowing our fitting routines to scale the model to the mean level of the ``pseudo-continuum'' would underestimate the true continuum level.  To address this problem, we assume that, at high signal-to-noise ratios, small dips in the spectrum are not noise features, but weak absorption lines, and we normalize the spectrum such that its small-scale features peak at a value of 1.0.  To estimate the uncertainty inherent in this technique, we perform each fit twice, once with the model continuum fixed at unity and again with the model scaled by a factor of 0.97, which brings its continuum closer to the pseudo-continuum of the observed spectrum.  An example is presented in \figref{fig_carbon3}.  The difference in the two abundances is an estimate of the systematic error in our abundance estimates. We add this term and the statistical error in quadrature to compute our final error. 

Notes on individual elements:

{\em Magnesium:}\/ In the COS spectrum near the \ion{Mg}{2} $\lambda \lambda1734.9, 1737.6$ doublet, the continuum is poorly constrained.

{\em Silicon:}\/   The \sithree\ 4552.6 \AA\ feature is much stronger than the \ntwo\  4552.5 \AA\ line with which it is blended, so we include it in our estimate of the silicon abundance.

{\em Phosphorus:}\/ \ion{P}{3} has resonance lines at 998.0 and 1003.6 \AA, but they lie among a forest of iron and other lines that are not included in our simple models.  To limit the effects of blending on the wings of these \ion{P}{3} features, we fit only the line cores, a spectral region roughly the width of a single resolution element.

{\em Sulfur:}\/  To limit the effects of blending, we fit only the cores of the sulfur lines in the \fuse\/ spectrum.

{\em Chlorine:}\/  The \ion{Cl}{3} resonance lines at 1005.3, 1008.8, and 1015.0 \AA\ lie in a crowded region of the spectrum, so we fit only the line cores and only the 1005.3 and 1008.8 \AA\ lines.  

{\em Argon:}\/ In the \fuse\/ spectrum near the \ion{Ar}{3} $\lambda 1002$ line, the continuum level is poorly constrained.

{\em Chromium:}\/  There are many \ion{Cr}{3} lines between 1032 and 1042 \AA.  Most are blended with other species, but the region between 1039 and 1041 \AA\ contains a half dozen lines that are reasonably well isolated.

{\em Nickel:}\/ In the \fuse\/ spectrum near the \ion{Ni}{3} $\lambda 980$ line, the continuum level is poorly constrained.

\section{Discussion}\label{sec_discussion}

\subsection{Model Atmospheres}\label{sec_models_discussion}

\subsubsection{Hydrogen}

Our models successfully reproduce the H$\alpha$ line profile, which has a distinctive shape near the line center.  To do so, we call TLUSTY with the parameter ICOLHN = 1 (the default value), which triggers the use of hydrogen collisional excitation rates taken from \citet{Przybilla:Butler:2004}.  Older versions of TLUSTY (prior to version 202), which employ the collisional rates of \citet{Mihalas:1975}, cannot reproduce this feature.  In \figref{fig_hydrogen}, we plot the best-fit spectra generated with TLUSTY versions 200 (blue) and 205 (red); the red curve better reproduces the line profile.  SYNSPEC version 51 offers several sets of hydrogen line profiles.  We have experimented with both the \citet{Lemke:1997} and \citet{Tremblay:Bergeron:2009} profiles and find that the resulting spectra are indistinguishable.

\subsubsection{Helium}

\begin{figure}
\plotone{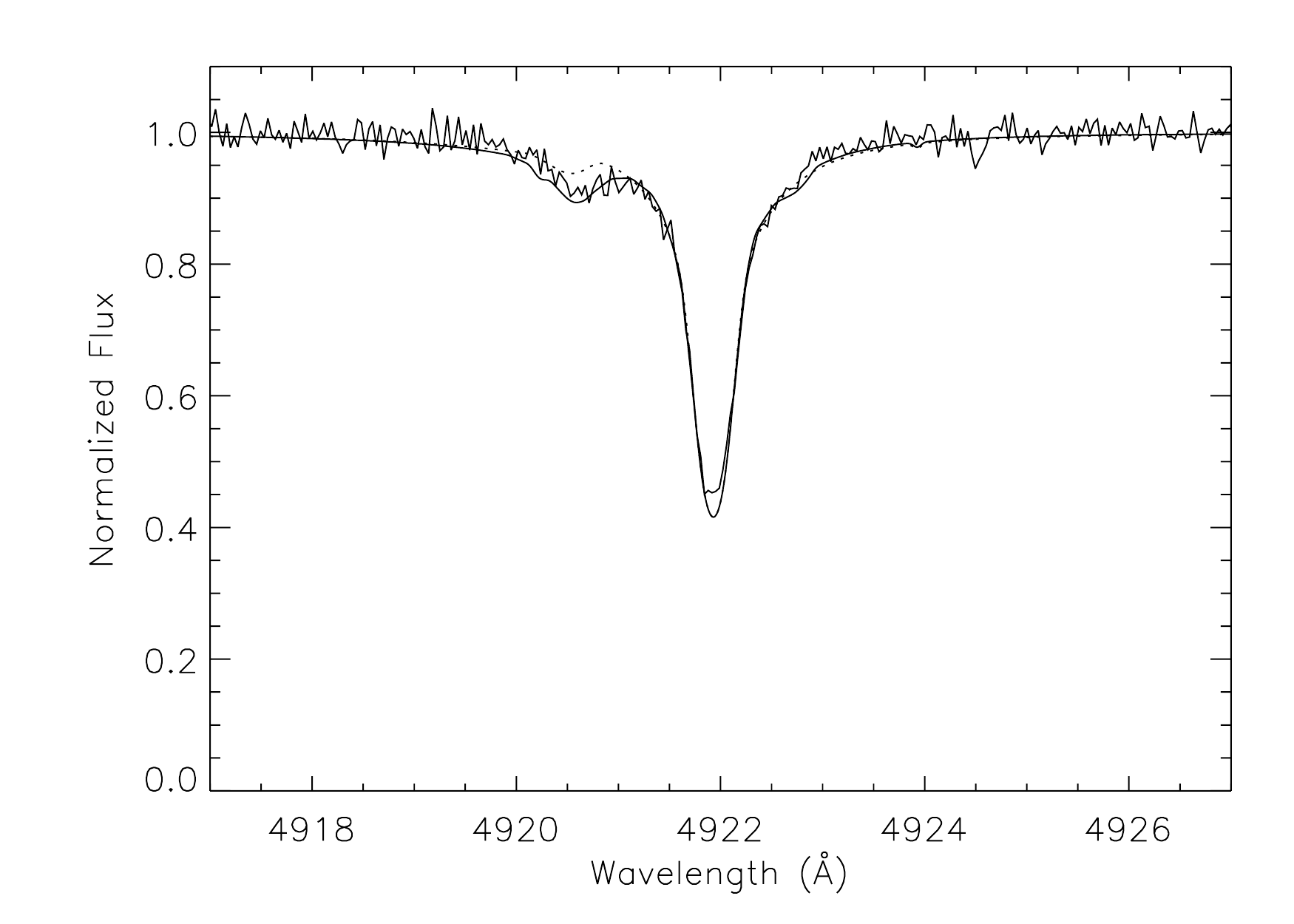}
\caption{\heone\ $\lambda 4922$ feature in the optical spectrum of Barnard~29 with synthetic spectra generated using line profiles from \citealt{Shamey:1969} (dotted line) and \citealt{BCS:1975} (solid line).}
\label{fig_better_he}
\end{figure}

Careful study of \figref{fig_helium} reveals weak forbidden components in the blue wings of the \ion{He}{1} $\lambda 4388$ and $\lambda 4922$ features that are not well reproduced by our model, while a similar component of the \ion{He}{1} $\lambda 4471$ feature is well fit. In SYNSPEC, the \ion{He}{1} $\lambda 4471$ profile is taken from \citet*[][hereafter BCS74]{BCS:1974}, who provide a set of line profiles for electron densities between $10^{13}$ and $10^{16}$ cm$^{-3}$. The profiles of the \ion{He}{1} $\lambda 4026$, $\lambda 4388$, and $\lambda 4922$ lines are from \citet{Shamey:1969}, who gives profiles for electron densities ranging from $10^{14}$ to $3\times10^{17}$ cm$^{-3}$.  The line-forming region of our model atmosphere extends to stellar radii with electron densities lower than $10^{14}$ cm$^{-3}$, a region of parameter space not included in the Shamey models.  Line profiles for the \ion{He}{1} $\lambda 4922$ line were computed by \citet*[][hereafter BCS75]{BCS:1975}. Their calculations are based on the work of BCS74 and cover the same range of electron densities.  We have incorporated the BCS75 profiles into SYNSPEC.
As shown in \figref{fig_better_he}, the absorption feature at 4920.5 \AA\ is now well reproduced.  Unfortunately, line profiles for the \ion{He}{1} $\lambda 4388$ line at electron densities lower than $10^{14}$ cm$^{-3}$ are not available, preventing us from computing a satisfactory model for this line.

\begin{figure}
\plotone{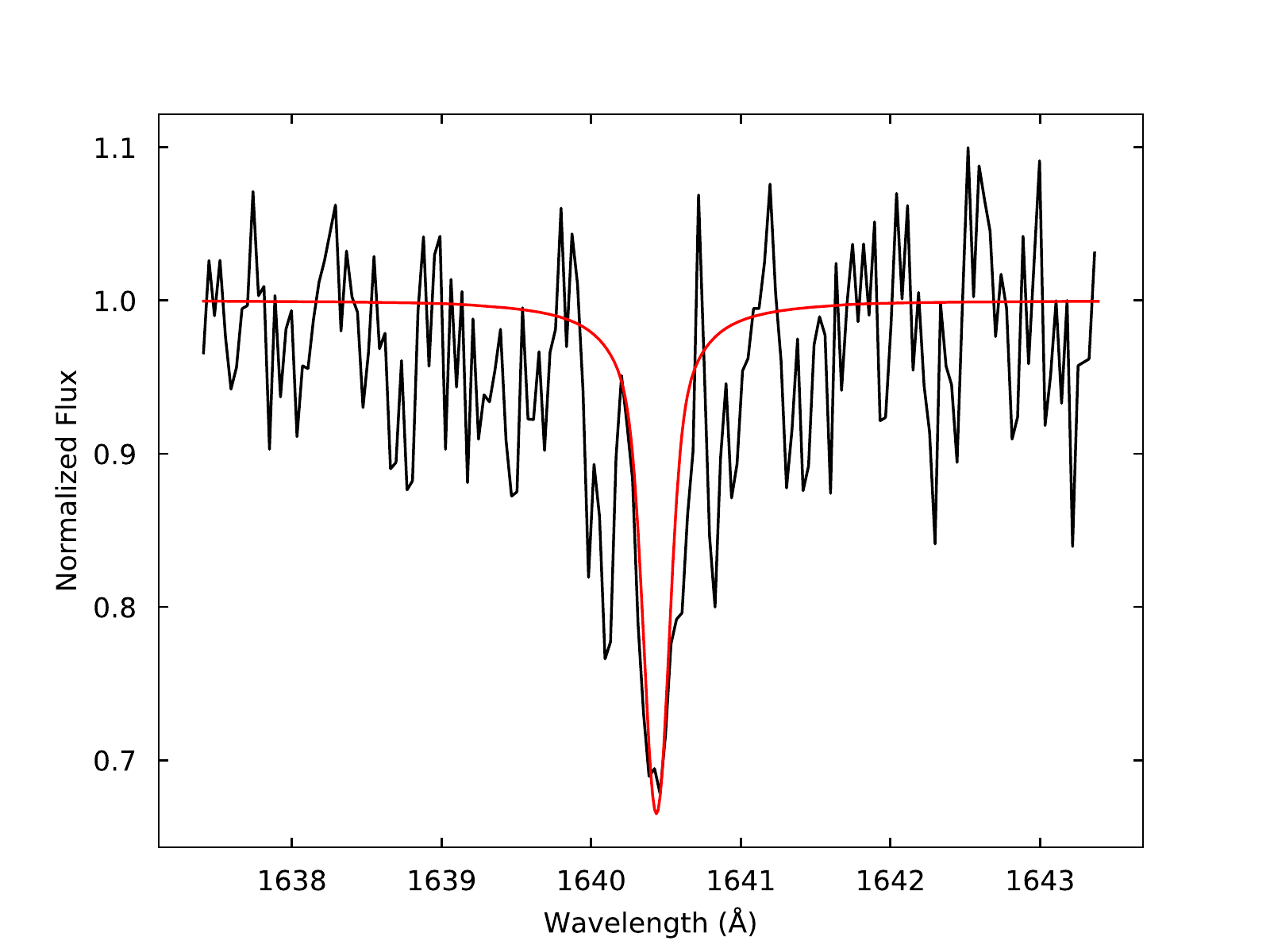}
\caption{The \hetwo\ $\lambda 1640$ feature in the COS spectrum of Barnard 29.   The red curve is a model with the helium abundance derived from our optical data.  For this figure, the spectrum has been normalized, binned by three pixels, and shifted to a laboratory wavelength scale.}
\label{fig_he_1640}
\end{figure}

The only \hetwo\ feature in our data is at 1640.4 \AA.  We do not use this line when deriving the effective temperature or helium abundance, because the spectrum is rather noisy in this region and the \hetwo\ line is blended with several metal features.  Instead, we simply plot in \figref{fig_he_1640} the COS spectrum (black curve) and a model with the helium abundance derived from our optical data (red curve).  The \hetwo\ profile is well reproduced by this model.

\subsubsection{Silicon II}\label{sec_silicon2}

\begin{figure}
\plotone{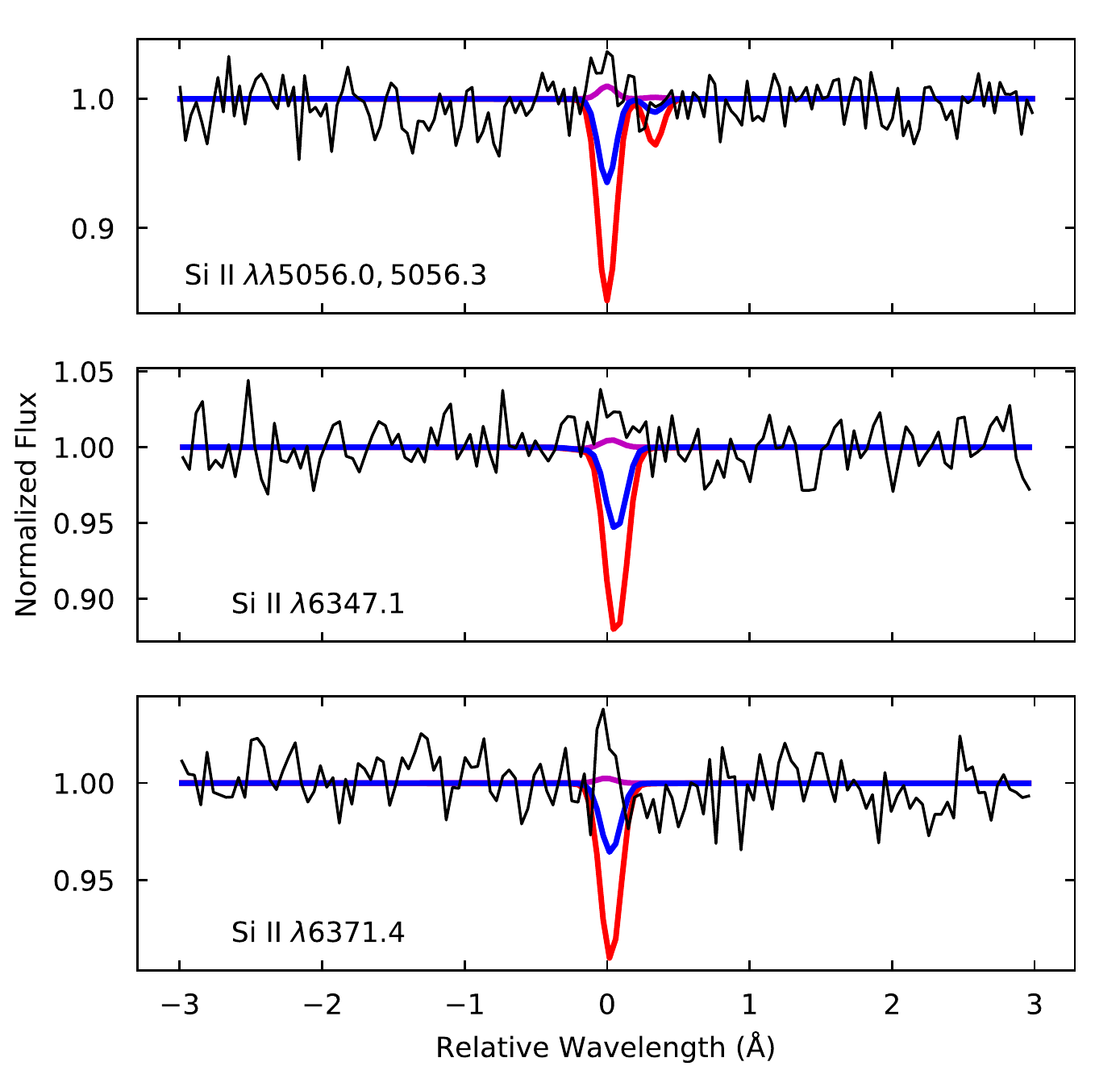}
\caption{\sitwo\ emission in the HIRES spectrum of Barnard 29.   The red curve represents our initial fit to the observed \ion{Si}{3} features, which yields \abund{Si} = $-5.12$.  At this abundance, the \ion{Si}{2} lines are predicted to be strong absorption features.  The blue curve represents a second model, using more sophisticated silicon model atoms, also fit to the \ion{Si}{3} features.  With an abundance of \abund{Si} = $-5.57$, this model predicts weaker \ion{Si}{2} absorption.  If we further reduce the abundance, we find that these lines go into emission at abundances lower than \abund{Si} = $-5.80$; the emission reaches a maximum around \abund{Si} = $-6.4$ (magenta curve).}
\label{fig_hires_sitwo}
\end{figure}

Our initial set of model stellar atmospheres employed the silicon model atoms that \citet{Lanz:Hubeny:2003, Lanz:Hubeny:2007} built to compute their grids of NLTE line-blanketed model atmospheres of B and O stars.  The resulting synthetic spectra reproduce the \sithree\ lines in Table \ref{tab:lines_hires}, but they also predict strong absorption from the \sitwo\ features at 5041.0, 5056.0, 5056.3, 6347.1, and 6371.4 \AA.  Only the 5056.0, 6347.1, and 6371.4 \AA\ lines are observed, and all three are in emission (\figref{fig_hires_sitwo}).  In an attempt to alleviate this discrepancy, we employ a more elaborate set of silicon model atoms (retrieved from \href{http://aegis.as.arizona.edu/~hubeny/pub/}{I.~Hubeny's website}) that include additional energy levels and allowed transitions: The new \sitwo\ model contains 70 individual energy levels and considers 940 allowed transitions, the \sithree\ model contains 122 levels and considers 1648 transitions, and the \sifour\ model contains 53 levels and considers 760 transitions.  We use these model atoms to generate a new grid of model stellar atmospheres.  Fits of the resulting synthetic spectra to the \sithree\ lines yield a lower silicon abundance and weaker \sitwo\ absorption (blue curves in \figref{fig_hires_sitwo}).  The models do predict faint \sitwo\ emission at still lower abundance levels (magenta curves), but the lines weaken as the silicon abundance is further decreased.  While our models do not produce \sitwo\ emission lines strong enough to match the observed emission, they do explain qualitatively the behavior of the lines by predicting the \sitwo\ $\lambda\lambda$5056, 6347, and 6371 lines in emission and the \sitwo\ $\lambda$4128.0 line in absorption, as reported by \citet{Thompson:07}.

Observations of weak metallic emission lines in the optical spectra of B-type stars have been reported by several authors (\eg, \citealt{Sigut:2000} and \citealt{Wahlgren:2000, Wahlgren:2004}). In particular, \citet{Sadakane:2017} reported the detection of weak \sitwo\ $\lambda$6239 and \altwo\ $\lambda$6237 emission in the optical spectra of B-type stars with low rotational velocities. Several of these stars have effective temperatures similar to that of Barnard 29, but their gravities are higher by 0.6 to 1.1 dex. They are also significantly more metal rich. The \sitwo\ $\lambda$6239 line is not observed in the spectrum of Barnard~29 because of the star's low silicon abundance; however, by raising the silicon abundance of our models to \abund{Si} = $-4.80$, we can generate synthetic spectra with the \sitwo\ $\lambda$6239 line in emission and the \sitwo\ $\lambda\lambda 5056.0,$ 6347.1, and 6371.4 lines in absorption, reproducing the pattern seen by \citeauthor{Sadakane:2017}. This result strengthens our confidence in these models and demonstrates that variations in the silicon abundance can explain, at least qualitatively, the pattern in the \sitwo\ lines that is observed in both Barnard~29 and some of the B-type stars analyzed by \citeauthor{Sadakane:2017}.

\subsubsection{Silicon III}\label{sec_silicon3}

\begin{figure}
\plotone{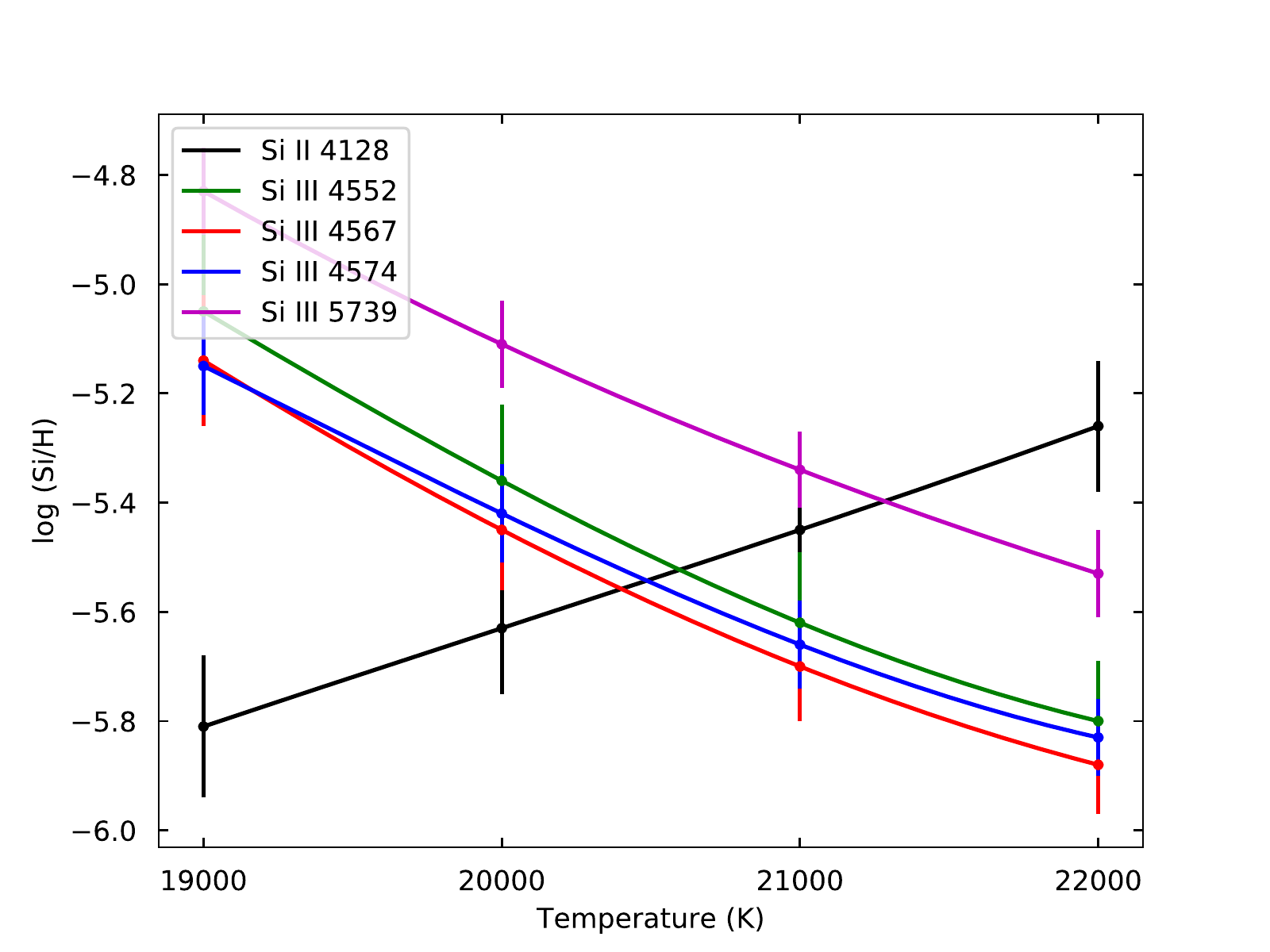}
\caption{Deriving the effective temperature from silicon lines in the optical spectrum of Barnard~29.  The \sitwo\ curve is computed from the equivalent width of  \sitwo\ $\lambda 4128$  published by \citealt{Thompson:07}.  The \sithree\ curves are derived from features in our Keck spectrum.  Points with error bars represent the abundance derived from model fits to each absorption feature.  Solid lines are low-order polynomial fits to the measured points.}
\label{fig_thompson}
\end{figure}

 \citet{Conlon:94} derived an effective temperature of \teff\ $= 20,000 \pm 1000$ K by requiring that the \sitwo\ line at 4128 \AA\ and the \sithree\ triplet at 4552, 4567, and 4574 \AA\ yield consistent silicon abundances.  \citet{Thompson:07} employed the same technique (with the addition of \sithree\ $\lambda 5739$) and achieved the same result.  Our equilibrium analysis includes these \sithree\ lines, yet yields an effective temperature of 21,400 K.  Why might that be?  In \figref{fig_thompson} we plot the abundances derived from each of the \sithree\ features in our Keck spectrum.  We cannot model our \sitwo\ lines, so we use the equivalent width of the \sitwo\ $\lambda 4128$ line published by \citeauthor{Thompson:07} to derive a silicon abundance.  We see that the \sitwo\ curve crosses those for the \sithree\ triplet at \teff\ $\sim 20,500$ K, while it crosses the curve for \sithree\ $\lambda 5739$ at \teff\ $\sim 21,300$ K.  This plot is derived from models using oscillator strengths from \citet{Kurucz:92}; models using oscillator strengths from the more recent compliation of \citet{Kelleher:Podobedova:2008} show an even greater difference ($\sim 1000$ K) between the intersection points.  Apparently, the \sithree\ triplet yields systematically lower abundances than does \sithree\ $\lambda 5739$, leading to a lower effective temperature from an equilibrium analysis.  The Keck \sithree\ curve in \figref{fig_temp_abundance} represents the mean of the four \sithree\ curves shown here.

\subsubsection{Microturbulence}\label{sec_microturbulence}

Microturbulence is an {\em ad hoc} parameter originally introduced to reconcile discrepancies between theoretical and empirical curves of growth \citep{Mihalas:78}.  In a traditional abundance analysis based on absorption-line equivalent widths, the microturbulent velocity $\xi$ is adjusted to remove any trend in abundance for individual atomic transitions as a function of the lower excitation potential or line strength.  In practical terms, microturbulence corrects for (or obscures) deficiencies in plane-parallel atmospheric models, including errors in the treatment of collisional broadening, dynamical effects, inhomogeneities, and non-LTE effects.

\begin{figure}
\plotone{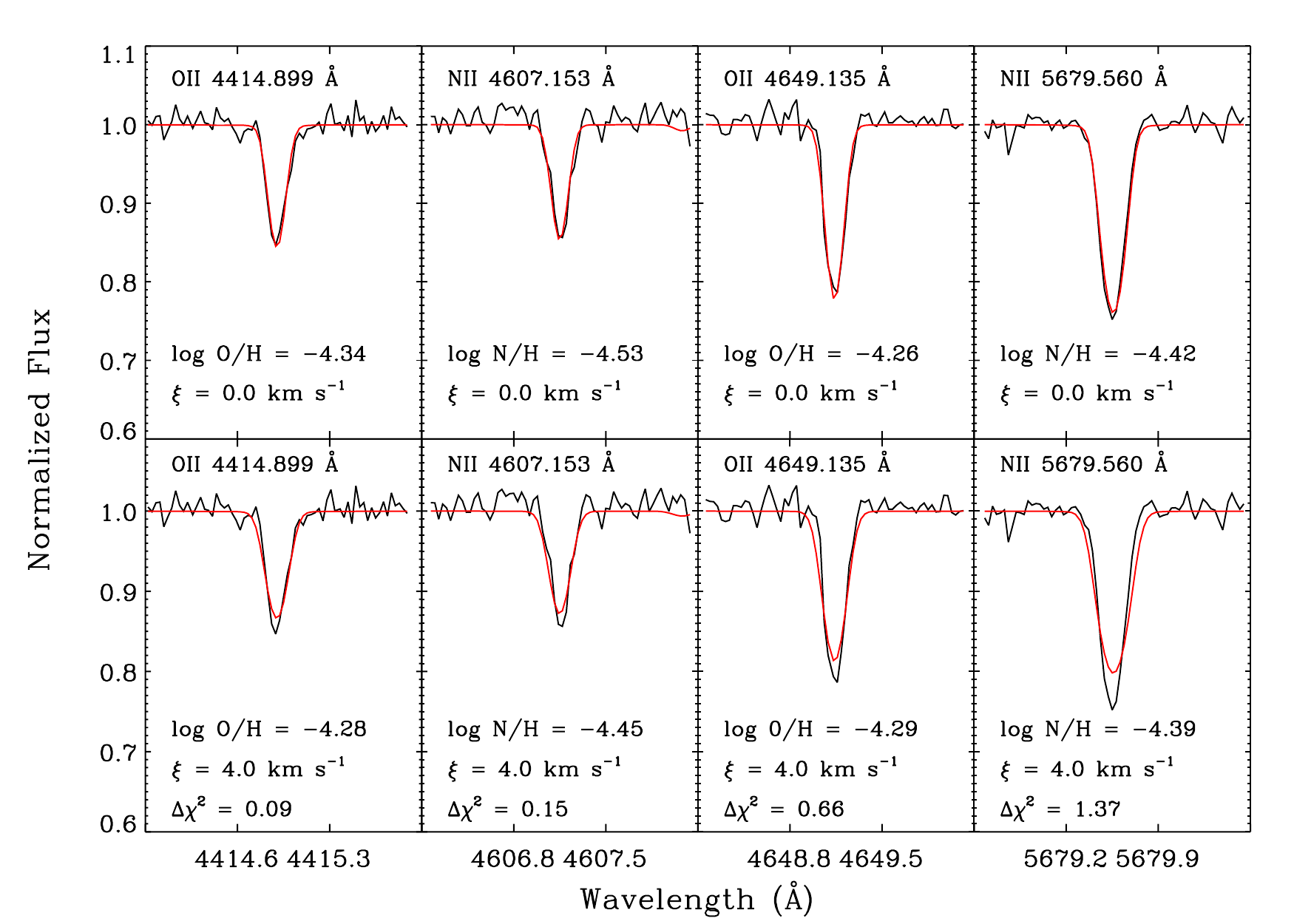}
\caption{{\em Top row:}  Nitrogen and oxygen features from the Keck spectrum of Barnard~29 fit using models with $\xi = 0$ \kms.  {\em Bottom row:} The same features fit using models with $\xi = 4$ \kms.}
\label{fig_vturb}
\end{figure}

In calculating our atmospheric models, we have set $\xi = 0$ \kms.  To explore the implications of this decision, we calculate a second set of models assuming $\xi = 4$ \kms, the value adopted by \citet{Thompson:07}, and compare them to the data.  The top row of \figref{fig_vturb} shows a set of nitrogen and oxygen features fit to the Keck spectrum using models with $\xi = 0$ \kms.  For this figure, each line is fit independently, and the resulting abundance is listed.  The bottom row shows the same features fit using models with $\xi = 4$ \kms.  While the abundances are little changed, the quality of the fit is obviously poorer, as evidenced by the increase in $\chi^2$.  We conclude that the data are best reproduced by models with $\xi = 0$ \kms.

\subsection{Photospheric Abundances}\label{sec_abundance_discussion}

Galactic globular clusters host multiple stellar populations. First-generation (FG) stars display abundances typical of halo field stars, while second-generation (SG) stars, which may have multiple subpopulations, are enriched in Na and Al and depleted in O and Mg.  Models suggest that the second generation is formed from gas polluted by material expelled by massive stars of the first generation.  (For details, see the review by \citealt{Bastian:Lardo:2018}.)  \citet{Meszaros:2015} identified two stellar populations on the RGB of M13. FG stars (blue lines in \figref{fig_abundance}) are richer in O and poorer in Al than SG stars (green lines).  Barnard~29 is clearly a FG star, richer in O and poorer in Al.  The star is depleted in C and enhanced in N; otherwise, its abundances appear to have changed little since it left the RGB.  In particular, its low carbon abundance ($N_C/N_O=0.03$) indicates that the star did not undergo third dredge-up.

In addition to the abundance trends attributable to multiple populations, \citet{Meszaros:2015} find a clear correlation in the abundance of carbon with effective temperature in M13.  The authors interpret this trend as a sign of ``deep mixing,'' a nonconvective mixing process that results in a steady depletion of the surface carbon abundance and an enhancement in the nitrogen abundance in low-mass stars as they evolve up the RGB \citep{Gratton:2004}.  The abundance pattern seen in Barnard~29 --- C depletion and N enhancement relative to the average value for FG stars on the RGB --- suggests that the star experienced the full impact of deep mixing.

\subsubsection{Iron Abundance}\label{sec_iron}

The iron abundance of Barnard~29 has been a puzzlement for more than two decades.  \citet{Conlon:94} set an upper limit of \abund{Fe} $< -5.30 \pm 0.30$ from an optical spectrum of the star.  \citet{Dixon:Hurwitz:98} observed Barnard~29 with the Berkeley Spectrometer, which spans the same wavelength range as \fuse, but at lower resolution.  Fitting spectra derived from \citet{Kurucz:92} models to the region between 1115 and 1160 \AA, they derived \abund{Fe} = $-6.70^{+0.22}_{-0.26}$.  Both \citet{Moehler:98} and \citet{Thompson:07} modeled the star's GHRS spectrum, deriving \abund{Fe} = $-6.79 \pm 0.10$ and \abund{Fe} = $-6.6 \pm 0.1$, respectively.  All of these FUV-derived values are significantly lower than the cluster iron abundance of \abund{Fe} = $-6.05$ \citep{Meszaros:2015}.  Strangely, \citeauthor{Thompson:07} found that fits to the star's optical spectrum yield \abund{Fe} $-5.93 \pm 0.12$, a value consistent with the cluster metallicity.  Canonical stellar-evolution theory   predicts no evolutionary changes in the iron abundance of low-mass stars like those in Galactic globular clusters today \citep{Iben:Renzini:83}.

We have independently derived the iron abundance of Barnard~29 from its Keck, \fuse, COS, and GHRS spectra.  Our values are  consistent with one another (\figref{fig_only_abundance}) and with the cluster iron abundance (\figref{fig_abundance}).  To understand the difference between our results and those of previous authors, we have conducted a series of numerical experiments, systematically testing the assumptions made by \citet{Moehler:98}.  Following them, we fit the entire suite of \fethree\ features in the GHRS spectrum.  Using models with a microturbulent velocity $\xi = 0$ \kms, we derive an iron abundance \abund{Fe} = $-6.16 \pm 0.16$.  \citeauthor{Moehler:98} adopt the value $\xi = 10$ \kms\ derived by \citet{Conlon:94}.  Using models with $\xi = 10$ \kms, we derive an iron abundance \abund{Fe} = $-6.79 \pm 0.09$, which matches the \citeauthor{Moehler:98} result.  Assuming a spectral resolution of FHWM = 0.07 \AA, employing LTE models, or adopting the atomic data presented by \citeauthor{Moehler:98} does not further reduce the derived iron abundance. 

Abundances derived from FUV spectra are sensitive to the assumed value of the microturbulent velocity.  Many FUV lines are saturated, and over-estimating the microturbulence results in an under-estimate of the abundance. \citet{Conlon:94} derive $\xi = 10 \pm 2$ \kms\ by minimizing the scatter in the abundance derived from 17 oxygen lines in the star's optical spectrum.  \citet{Dixon:Hurwitz:98} employ Kurucz models with $\xi = 2$ \kms, while \citet{Thompson:07} derive $\xi = 4 \pm 2$ \kms\ from fits to \otwo\ and \sithree\ lines.  

%

\subsubsection{Zirconium Abundance}\label{sec_zirconium}

In a preliminary analysis of the \fuse\/ and COS data \citep{Dixon:11}, we attributed the strong absorption feature at 1184 \AA\ to \ion{Zr}{4}; however, recent calculations of \ion{Zr}{4} oscillator strengths by \citet{Rauch:2017} raise doubts about this identification.  \ion{Zr}{4} $\lambda 1184$ should be accompanied by \ion{Zr}{4} $\lambda 1220$, as both are resonance transitions, yet the $\lambda 1220$ feature is not seen.  The hot sub-dwarf B star CPD$-64$~481 has an effective temperature \teff\ = 27,500 K \citep{OToole:2005}, similar to that of Barnard~29.  Examination of a high-resolution spectrum of CPD$-64$~481 obtained with the \hst\/ STIS E140H grating reveals that \ion{Zr}{4} $\lambda 1184$ is blended with another, unidentified feature.  We conclude that the 1184 \AA\ feature cannot be used to derive the Zr abundance of stars observed at the resolution of \fuse\/ or COS.


\subsection{Stellar Mass and Luminosity}\label{sec_mass}

We can derive a star's radius, and from this its mass and luminosity, by comparing its observed and predicted flux.  The spectral irradiance of Barnard~29 has been measured in several bands:  $B = 12.934$, $V = 13.116$, $I = 13.303$ mag \citep{Sandquist:2010}, for which we assume an uncertainty of 0.003 mag.  The extinction toward M13 is \ebv\ = $0.02 \pm 0.01$ mag \citep{Harris:2010}.  We model the stellar continuum using an NLTE model, similar to the H+He+N+O+Fe models described in Section \ref{sec_models} but with our best-fit atmospheric parameters, scaled by a \citet{CCM:89} extinction curve with $R_V = 3.1$.  We compute synthetic stellar magnitudes using the Python package Pysynphot \citep{Lim:2015}, adjusting the magnitude zero points as described by \citet{Dixon:2017}.  The data are best fit with a scale factor $\phi = (1.216 \pm 0.002) \times 10^{-21}$.  We repeat the fit assuming an extinction \ebv\ = 0.03 and get $\phi = (1.251 \pm 0.002) \times 10^{-21}$.  Combining this change with the statistical error yields a final scale factor $\phi = (1.22 \pm 0.04) \times 10^{-21}$.  

In the synthetic spectra generated by SYNSPEC, the flux is expressed in terms of the flux moment, $H_\lambda$.  If the star's radius and distance are known, then the scale factor required to convert the model to the flux at earth is $\phi = 4 \pi (R_* / d)^2$ \citep{Kurucz:79}.  Unfortunately, we cannot use results from {\em Gaia}\/ Data Release 2 to constrain the distance to M13, as {\em Gaia}\/ globular-cluster parallaxes suffer from significant systematic errors \citep{Helmi:2018}.  Recent published values range from 7.1 kpc \citep{Harris:2010} to 7.9 kpc \citep{O'Malley:2017}.  Adopting this distance range and our scale factor, we derive a stellar radius  $R_*/R_{\sun}$ between 3.11 and 3.45.  Applying our best-fit surface gravity (\logg\ = 3.1), we find that the stellar mass $M_*/M_{\sun}$ is between 0.45 and 0.55.  Finally, combining the stellar radius with our best-fit effective temperature (\teff\ = 21,400 K), we derive a stellar luminosity $\log L_*/L_{\sun}$ between 3.26 and 3.35, slightly greater than the range $\log L_*/L_{\sun}$ = 3.08 -- 3.25 estimated by \citet{Conlon:94}.  All derived values (radius, mass, and luminosity) scale with the cluster distance.


\subsection{Evolutionary Status}\label{sec_evolution}

\begin{figure}
\plotone{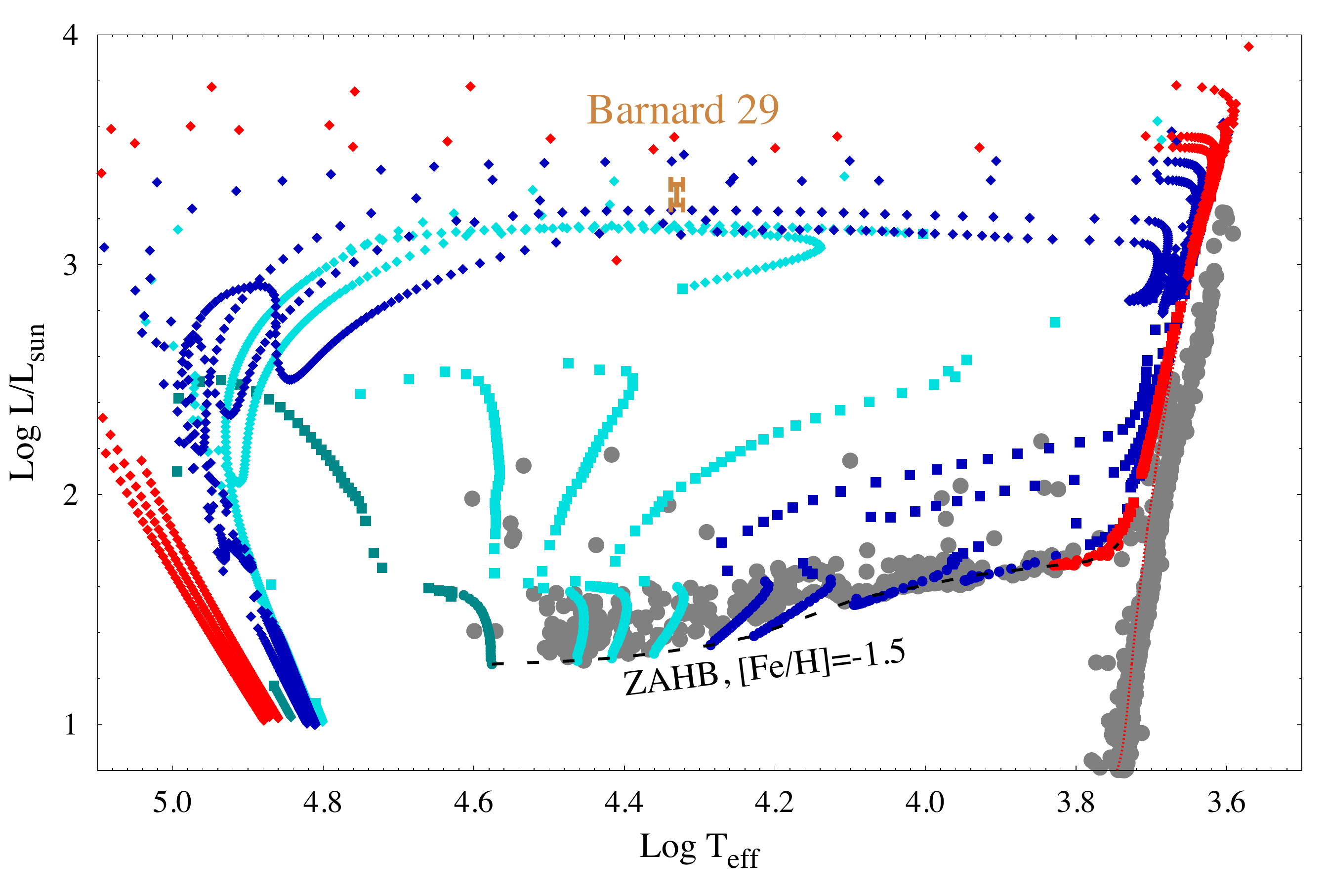}
\caption{Evolutionary tracks for stars similar to those of M13 during and after the horizontal-branch stage. Colors indicate stars that spend their core helium burning stage as Red Horizontal Branch stars (red), Blue Horizontal Branch stars (blue), and Extreme Horizontal Branch Stars (cyan). The hottest track (dark cyan) represents a Late Hot-Flasher sequence.  Circles are plotted every 10 Myr, squares are plotted every 1 Myr, and diamonds indicate intervals of  20,000 yr.  The vertical orange bar marks the effective temperature and range of allowed luminosities for Barnard~29.}
\label{fig_tracks}
\end{figure}

The post-HB evolution of M13 stars is illustrated in \figref{fig_tracks}.  The coolest (\teff\ $\lesssim$ 8,000 K), most massive ($M_{\rm ZAHB} \gtrsim 0.70 M_{\sun}$) stars populate the red horizontal branch (RHB, red points).  As they depart the HB, they evolve to the red and ascend the AGB.  Stars on the blue horizontal branch (BHB, blue points) are hotter (8,000 K $\lesssim$ \teff\ $\lesssim$ 20,000 K) and have lower masses ($0.52 \lesssim M_{\rm ZAHB} \lesssim 0.70M_{\sun}$).  Most post-BHB stars also climb the AGB and reach the thermal-pulsing AGB phase, but only for a short time.  The hottest (\teff\ $\gtrsim$ 20,000 K), least-massive ($M_{\rm ZAHB} \lesssim 0.52M_{\sun}$) stars populate the extreme horizontal branch (EHB, cyan points).  Most of these stars do not climb the AGB after the He-core burning phase, but evolve to high luminosities with little change in temperature.  Post-EHB stars are also known as AGB-manqu\'{e} stars.  Stars near the boundary between the EHB and the BHB (\teff\ $\sim$ 20,000 K, $M_{\rm ZAHB} \sim 0.52 M_{\sun}$) follow an intermediate path: they climb the AGB, but depart before reaching the thermal-pulsing phase, becoming post-early AGB (post-EAGB) stars.

The tracks plotted in \figref{fig_tracks} represent an extension of
the work presented by \citet{Miller_Bertolami:2016}.  The models are
computed for [Fe/H] = $-1.5$ and a zero-age main-sequence (ZAMS) mass
of $M_{\rm ZAMS}=0.83$ \msun\ (age 11.7 Gyr), assuming a scaled-solar
metal content with initial abundances $Z_{\rm ZAMS} = 0.000548$, $Y_{\rm
ZAMS} = 0.246096$, and $X_{\rm ZAMS} = 0.753356$.  Winds on the RGB are
adjusted to populate the extreme, blue, and red horizontal branches.
Zero-age horizontal-branch (ZAHB) masses are $M_{\rm ZAHB}= 0.83$,
0.75 and 0.70 \msun\ (final masses $M_{\rm WD}=0.557$, 0.550, and
0.540\msun) for the RHB (red points); $M_{\rm ZAHB}= 0.65$, 0.60,
0.55, and 0.53 \msun\ (final masses $M_{\rm WD}=0.525$, 0.519, 0.505,
and 0.501 \msun) for the BHB (blue points); and $M_{\rm ZAHB}= 0.51$,
0.50, and 0.495 \msun\ (final masses $M_{\rm WD}=0.500$, 0.499, and
0.495 \msun) for the EHB (cyan points).
The dark-cyan points show the evolution of the hottest possible post-EHB
model. This sequence corresponds to a Late Hot-Flasher sequence (see
\citealt{Battich:2018} for a detailed explanation) that
undergoes a violent H-burning event during the He-core flash and ends as a
H-deficient ZAHB model (surface abundances [H, He, C, N] = [$2.3\times 10^{-4}$,
0.9636, 0.024, 0.012] by mass fraction), with a ZAHB mass of $M_{\rm ZAHB}=0.49$ \msun\
(final mass $M_{\rm WD}=0.489$ \msun).

To these evolutionary tracks we have added a sample of M13 stars from \citet{Nardiello:2018}, who provide a five-band catalog of stellar magnitudes from the  \hst\/ UV Legacy Survey of Galactic Globular Clusters \citep{Piotto:2015} and the ACS Globular Cluster Survey \citep{Sarajedini:2007}.  Effective temperatures are estimated from F275W $-$ F438W colors, and luminosities from F606W magnitudes.  We use \citet{Castelli:Kurucz:03} models with [Fe/H] = $-1.5$ to determine these relationships.  The cluster distance  and reddening are taken from \citet{Harris:2010}.  Barnard~29 is indicated by a vertical orange bar that spans the range of luminosities derived in Section \ref{sec_mass}.

M13 hosts only a handful of RHB stars, but its BHB and EHB are well populated.  In
\figref{fig_tracks}, we see a number of hot stars more luminous than the HB.  Some
appear to be following the post-BHB tracks toward the AGB, and some the post-EHB
tracks toward the white-dwarf cooling curve.  \citet{Sandquist:2010} estimate that
roughly one-third of the post-HB stars in M13 are on post-EHB tracks.

The derived mass and luminosity of Barnard~29 depend on its distance.  
If M13 lies at the low end of the allowed range ($d \sim 7.1$ kpc), then $M_*/M_{\sun} \sim 0.45$ and $\log L_*/L_{\sun} \sim 3.26$.
These values have interesting implications for the star's evolutionary history.   
No HB star can have a mass less than 0.49 \msun, 
because core helium ignition is impossible at such low masses.  An
evolved star with $M_*/M_{\sun} < 0.49$ must have experienced extreme
mass loss on the RGB (more even than the cluster's EHB stars), left
the RGB before the helium-core flash, and is now evolving directly
from the RGB to the white-dwarf cooling curve, where it will
become a helium-core white dwarf.   
The fact that M13 has a large population of extreme EHB stars suggests that 
it may also have a population of lower-mass post-RGB stars, 
because a finely-tuned mass-loss process would be required to remove just enough
material from RGB stars to produce a large population of the hottest EHB stars
without creating any lower-mass objects.  
So it is not unthinkable that Barnard~29 is a post-RGB star.
On the other hand, only the most massive post-RGB stars would be as luminous as Barnard~29.  
According to stellar-evolution models, a post-RGB star with the luminosity of Barnard~29  
would have a mass $M_*/M_{\sun} = 0.48$, greater than the derived value for this distance.
Consequently, a post-RGB origin for Barnard 29 is unlikely.

If the cluster lies at the high end of the allowed range ($d \sim 7.9$ kpc), 
then $M_*/M_{\sun} \sim 0.55$ and $\log L_*/L_{\sun} \sim 3.35$, values consistent with the star's being a
post-BHB star.  At an intermediate distance, the mass of Barnard~29 would lie between these extremes;
values $M_*/M_{\sun} \lesssim 0.52$ would place the star on one of the post-EHB evolutionary tracks.  
We can exclude the hottest EHB track in \figref{fig_tracks}, which yields a white dwarf with mass around 0.495 \msun.
It traces the path of AGB-manqu\'{e} stars that evolve directly from the HB to the white-dwarf cooling curve.  
Barnard~29 is clearly cooler and more luminous than this track, suggesting a mass $M_*/M_{\sun} \gtrsim 0.5$.


In light of stellar-evolution models, it seems likely that Barnard~29 is a post-HB star evolving from a ZAHB star with $M_{\rm ZAHB}$ between 0.50 and 0.55 \msun, a range spanning the EHB/BHB boundary.  A post-HB star would have fully ascended the RGB, consistent with our suggestion that Barnard~29 experienced the full effects of nonconvective mixing on the RGB.  Its relatively low mass is consistent with our conclusion that the star did not experience third dredge-up.  

\section{Summary and Conclusions}\label{sec_conclusions}

We have performed a spectral analysis of the UV-bright star Barnard~29 in M13.  By requiring multiple ionization states of C, N, O, Si, and S to yield consistent abundances, we derive an effective temperature \teff\ = $21,400 \pm 400$ K.  We derive a surface gravity \logg\ = $3.10 \pm 0.03$ from the star's Balmer lines.  Using the latest version of TLUSTY, we are able to reproduce the observed H$\alpha$ profile.  By adding \heone\ line profiles to SYNSPEC, we reproduce the absorption feature at 4920.5 \AA.  Although our models predict faint \sitwo\ emission in the optical, they cannot reproduce the strength of the observed features.  We derive the photospheric abundances of He, C, N, O, Mg, Al, Si, P, S, Cl, Ar, Ti, Cr, Fe, Ni, and Ge.  Barnard~29 exhibits an abundance pattern typical of first-generation stars in M13, though an underabundance of C and an overabundance of N suggest that the star experienced nonconvective mixing on the RGB.  This pattern appears to have changed little since the star left the RGB.  In particular, the star did not undergo third dredge-up.  Previous workers have found that the star's FUV spectra yield an iron abundance about 0.5 dex lower than its optical spectrum.  The iron abundances derived from our Keck, \fuse, COS, and GHRS spectra are consistent with one another and with the cluster value.  We attribute the difference to our use of model atmospheres without microturbulence, which is ruled out by the quality of the fits to the optical \ntwo\ and \otwo\ lines.  Barnard~29 lies in a region of the temperature-luminosity plane that is traversed by both post-BHB and post-EHB evolutionary tracks.  Comparison with stellar-evolution models suggests that it evolved from a ZAHB star of mass $M_*/M_{\sun}$ between 0.50 and 0.55, close to the EHB/BHB boundary.

\acknowledgments

We wish to thank A.~Beauchamp and P.~Bergeron for helpful discussions of \ion{He}{1} line profiles and J.~Valenti for discussions of microturbulence.
P.~C.\ is supported by the Canadian Space Agency under a contract with NRC Herzberg Astronomy and Astrophysics.
This work has made use of 
NASA's Astrophysics Data System (ADS);
the NASA/IPAC Extragalactic Database (NED), which is operated by the Jet Propulsion Laboratory, California Institute of Technology, under contract with the National Aeronautics and Space Administration;
the Mikulski Archive for Space Telescopes (MAST), hosted at the Space Telescope Science Institute, which is operated by the Association of Universities for Research in Astronomy, Inc., under NASA contract NAS5-26555;  
and
the Keck Observatory Archive (KOA), which is operated by the W.\ M.\ Keck Observatory and the NASA Exoplanet Science Institute (NExScI), under contract with the National Aeronautics and Space Administration.
%
%
%
%
%
%
Publication of this work is supported by the STScI Director's Discretionary Research Fund.

%

\vspace{5mm}
\facilities{HST(COS, GHRS), FUSE, Keck:I (HIRES)}


\software{CalFUSE \citep{Dixon:07}, CALCOS (v3.0) \citep{Fox:2015}, TLUSTY \citep{Hubeny:Lanz:95}, SYNSPEC \citep{Hubeny:88}, Pysynphot \citep{Lim:2015}, specnorm.py (\url{http://python4esac.github.io/plotting/specnorm.html})}

\end{document}